\documentclass[journal]{vgtc}                % final (journal style)
\ifpdf%                                % if we use pdflatex
  \pdfoutput=1\relax                   % create PDFs from pdfLaTeX
  \usepackage{graphicx}                % allow us to embed graphics files
  \DeclareGraphicsExtensions{.pdf,.png,.jpg,.jpeg} % for pdflatex we expect .pdf, .png, or .jpg files
\else%                                 % else we use pure latex
  \usepackage{graphicx}                % allow us to embed graphics files
  \DeclareGraphicsExtensions{.eps}     % for pure latex we expect eps files
\fi%

\graphicspath{{figures/}{pictures/}{images/}{./}} % where to search for the images

\usepackage{microtype}                 % use micro-typography (slightly more compact, better to read)
\PassOptionsToPackage{warn}{textcomp}  % to address font issues with \textrightarrow
\usepackage{textcomp}                  % use better special symbols
\usepackage{mathptmx}                  % use matching math font
\usepackage{times}                     % we use Times as the main font
\usepackage{cite}                      % needed to automatically sort the references
\usepackage{tabu}                      % only used for the table example
\usepackage{booktabs}                  % only used for the table example
\usepackage[linesnumbered,ruled]{algorithm2e}
\SetCommentSty{mycommfont}

\DontPrintSemicolon
\usepackage[shortlabels]{enumitem}
\usepackage{color}
\SetKwRepeat{Do}{do}{while}%
\usepackage{amsmath,amssymb}
\newcommand{\subscript}[2]{$#1 #2$}
\setlist[enumerate]{labelindent=0.5em,labelsep=0.5em,leftmargin=*,parsep=0.1em,topsep=0.25em} %item间距
\usepackage{hyperref}
\usepackage{siunitx}
\usepackage{circledsteps}
\pgfkeys{/csteps/inner ysep=3pt}
\pgfkeys{/csteps/inner xsep=3pt}

\usepackage{makecell}

\usepackage{utfsym}
\usepackage{makecell}
\usepackage{multirow}
\usepackage{color}
\usepackage{caption} 
\definecolor{applegreen}{rgb}{0.55, 0.71, 0.0}
\definecolor{ao(english)}{rgb}{0.0, 0.5, 0.0}
\def\halfcheckmark{\checkmark\kern-1.1ex\raisebox{.7ex}{\rotatebox[origin=c]{125}{--}}}
\onlineid{1394}

\vgtccategory{Research}
\vgtcpapertype{algorithm/technique}

\title{Dual Space Coupling Model Guided Overlap-Free Scatterplot}

\author{Zeyu Li, Ruizhi Shi, Yan Liu, Shizhuo Long, Ziheng Guo, Shichao Jia, and Jiawan, Zhang, \textit{Senior Member, IEEE}}
\authorfooter{
\item
 Z. Li, R. Shi, Y. Liu, S. Long, Z. Guo, S. Jia, and J. Zhang are with College of Intelligence and Computing, Tianjin University. E-mail: \{lzytianda, shiruizhi, 3019244195, 3019244112, sygzh6, jsc, jwzhang\}@tju.edu.cn.
}

\shortauthortitle{Biv \MakeLowercase{\textit{et al.}}: Global Illumination for Fun and Profit}
\abstract{The overdraw problem of scatterplots seriously interferes with the visual tasks. Existing methods, such as data sampling, node dispersion, subspace mapping, and visual abstraction, cannot guarantee the correspondence and consistency between the data points that reflect the intrinsic original data distribution and the corresponding visual units that reveal the presented data distribution, thus failing to obtain an overlap-free scatterplot with unbiased and lossless data distribution. A dual space coupling model is proposed in this paper to represent the complex bilateral relationship between data space and visual space theoretically and analytically. Under the guidance of the model, an overlap-free scatterplot method is developed through integration of the following: a geometry-based data transformation algorithm, namely DistributionTranscriptor; an efficient spatial mutual exclusion guided view transformation algorithm, namely PolarPacking; an overlap-free oriented visual encoding configuration model and a radius adjustment tool, namely $f_{r_{draw}}$. Our method can ensure complete and accurate information transfer between the two spaces, maintaining consistency between the newly created scatterplot and the original data distribution on global and local features. Quantitative evaluation proves our remarkable progress on computational efficiency compared with the state-of-the-art methods. Three applications involving pattern enhancement, interaction improvement, and overdraw mitigation of trajectory visualization demonstrate the broad prospects of our method.} % end of abstract


\CCScatlist{ % not used in journal version
 \CCScat{K.6.1}{Management of Computing and Information Systems}%
{Project and People Management}{Life Cycle};
 \CCScat{K.7.m}{The Computing Profession}{Miscellaneous}{Ethics}
}

\teaser{
  \centering
  \includegraphics[width=\linewidth]{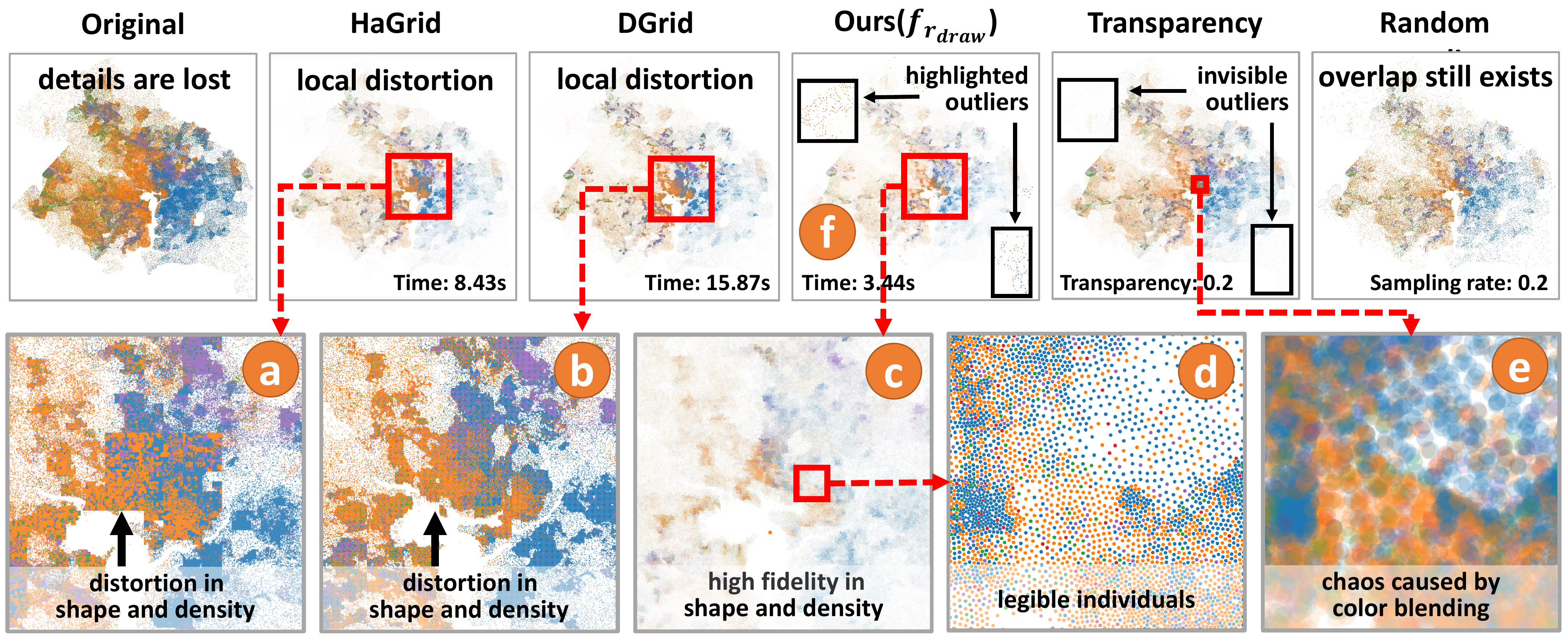}
  \caption{Results of our method and several representative methods on a dataset containing \num[group-separator={,}]{401950} data points and 5 categories. Sampling and transparency adjustment cannot fundamentally eliminate overdraw. The color blending induced by transparency seriously impedes the understanding of local details\Circled{e}. HaGrid and DGrid suffer from distortion in shape and density in high density regions\Circled{a}\Circled{b}. Our method can accurately characterize the density distribution globally and locally while highlighting outliers\Circled{c}\Circled{d}.}
  \label{fig:teaser}
}

\vgtcinsertpkg

\begin{document}
\setlength{\abovedisplayskip}{3pt}
\setlength{\belowdisplayskip}{3pt}


\firstsection{Introduction}

\maketitle
%% \section{Introduction} %for journal use above \firstsection{..} instead

For 2D scatterplot visualization, maintaining high-quality data distribution while avoiding overdraw is still an unsolved problem. 

Depending on the space in which the core operation is performed, existing solutions toward overdraw problem can be classified into three categories: data space, visual space, and hybrid methods. First, data space methods perform data transformation such as trimming, filtering, sampling, or aggregating operation, on the original data points to reduce the data volume. However, the asymmetrical correspondence between the data points and the visual units in visual space objectively introduces an endogenous contradiction between reducing overdraw and maintaining a lossless and unbiased data distribution. Second, visual space methods mainly focus on applying visual encoding adjustment and view transformation by elaborately configuring the size, position, transparency, or other visual channels of visual units. These methods can then be further classified into three sub-categories: node appearance adjustment\cite{wilkinson2012grammar}\cite{li2010model} whose strategy is to reduce the size and transparency of nodes; node dispersion\cite{dwyer2005fast-VPSC}\cite{gansner2010efficient-PRISM}\cite{nachmanson2016node-GTree}\cite{strobelt2012rolled-RWordle} which distributes nodes in an iteration process based on a physical or a mathematical optimization model; sub-space mapping \cite{hilasaca2019overlap-DGrid}\cite{cutura2021hagrid}\cite{fried2015isomatch} which injects the data nodes into a partition of visual space. However, adjusting the appearance of nodes cannot strictly avoid overlap, and the color blending caused by transparency leads to severe visual complexity. The two latter sub-categories may introduce serious distortions because they disregard the density preservation. Third, hybrid methods, such as bin aggregation\cite{heimerl2018visual-bin}\cite{beilschmidt2019efficient}\cite{luboschik2010new-color-weaving} and contour map\cite{collins2009bubble-contourmap}\cite{mayorga2013splatterplots}\cite{li2019galex}, relieve overdraw by replacing visual units that originally correspond to individual data points with visual objects, such as polygons and paths with higher level of visual abstraction. Abstraction leads to the loss of details. 

The two major drawbacks of the existing methods lie in the asymmetrical correspondence between the data points and visual units and the inconsistency between the original data distribution and the distribution presented in the scatterplot. To obtain a complete solution to scatterplot overdraw problem, first, the data-visual space mapping should be unbiased or lossless. No data points can be discarded, and all data points should correspond to a unique visual unit. Second, principles and guidelines should also be carefully developed to ensure the strict overlap elimination and safe interaction\footnote{no overlap occurs during the interaction} to handle the scalability issue of data and support interactive exploration in visual space. Third, as the fundamental goal, the final scatterplot should accurately reflect the original data distribution. The existing overlap removal methods also pursue similar objectives but they always seriously sacrifice one of them, leading to the realization of some goals but accompanied by serious negative effects.

In this paper, after re-examining the overdraw problem through a theoretical perspective, we regard it as an informal optimization problem under four formal criteria. In addition to emphasizing the safety of operations performed within a single space, a dual space coupling model is also proposed to represent the complex bilateral relationship between data and visual spaces analytically. Under the guidance of this model, we develop an overlap-free scatterplot visualization method comprising an unbiased and lossless geometry-based transcriptor that transcribes the data distribution into a set of discrete circles, an efficient circle packing algorithm that re-layouts these circles in visual space to ensure spatial mutual exclusion and reproduce the transcribed distribution to a tangible scatterplot, and a visual encoding configuration model to optimize the visual quality of the new scatterplot and ensure interaction safety. The proposed method can ensure complete and accurate information transfer between the data and visual spaces, maintaining consistency between the newly created scatterplot and the original data points on global and local features. Quantitative evaluations are conducted to compare with the state-of-the-art methods on time cost and five metrics that are designed for measuring the capability to preserve the features of original scatterplots. Three applications demonstrate the capability of our method to reveal the pattern hidden by overdraw, improve the efficiency of interactive exploration, and mitigate the overdraw problem in trajectory visualization.

The contributions of this paper can be summarized as follows:
\begin{itemize}
    \item We propose a dual-space coupling model to represent the complex relationship and design considerations within and between the data and visual spaces theoretically and analytically. The model introduces a new design space for promising overlap removal algorithms and interaction paradigms.
    \item We propose an overlap-free scatterplot method which integrates \textit{DistributionTranslator} (a geometry-based data transformation algorithm), \textit{PolarPacking} (an efficient circle packing algorithm), and a visual encoding configuration model.
    \item We develop an easy-to-use radius adjustment tool $f_{r_{draw}}$ on the basis of the configuration model to improve the visual quality of scatterplot and ensure interaction safety.
\end{itemize}

\section{related work}

% Depending on the space where the core operation takes place, existing methods to mitigate overdraw can be classified into the following three categories.
Under the topic of visual enhancement of scatterplots\cite{staib2016enhancing}\cite{chan2010flow}, this paper focuses on methods to mitigate overdraw.

\subsection{Data Space Methods}
Data space methods simply focus on data operation and completely ignore the stuff on visual units. Therefore, these methods simply pertain to the data transformation described in the classic visualization pipeline\cite{card1999readings}. Data reduction and jitter are two typical data space methods.

Data reduction methods alleviate overdraw by reducing data points, thus decreasing the visual units to be placed in visual space. Data sampling and aggregation are two commonly used data reduction strategies. Data sampling selects representative samples from the full set, while data aggregation aggregates subsets of the full set into newly created data points. However, they both suffer from inherent flaws on data loss, data bias, and visualization. The data loss is straightforward, while the data bias is caused by the unavoidable selection of data or/and goal. For example, data sampling methods have developed diverse sampling strategies but have all been designed for specific goals, such as maintaining relative density among regions \cite{bertini2004chance-density}\cite{chen2021pyramid}\cite{joia2015uncovering-svd}\cite{bertini2005improving}, emphasizing the spatial separation of samples \cite{chen2014visual-multi-class}\cite{wei2010multi-class-blue-noise}\cite{Farthest-point-sampling}, and preserving outliers\cite{liu2017visual-outlier}\cite{chen2019recursive}\cite{xiang2019interactive-outlier-bluenoise}. No one-size-fits-all sampling strategy exists. Hence, the choice of a goal/strategy causes bias. Data reduction methods pose a fundamental conflict between avoiding overdraw and preserving unbiased and lossless data distribution. Moreover, data reduction methods cannot eliminate overlaps because they completely ignore the size of visual units.

Typical jitter\cite{theus2008interactive-jittering} alleviates overdraw by randomly spreading data points in data space. However, jitter is unstable and cannot materially overcome overdraw. Instead, jitter may lose meaningful data features or even cause more serious overlap.

\subsection{Visual Space Methods}
In contrast to data space methods, visual space methods focus entirely on visual units in visual space, reducing overlap by optimizing their appearance or position. These methods ensure one-to-one correspondence between data points and visual units, avoiding information loss at the data level. Therefore, these methods come down to visual encoding and view transformation operation following the classic visualization pipeline\cite{card1999readings}. Appearance adjustment, node dispersion, and subspace mapping are three common visual space methods.

Appearance adjustment reduces overdraw by decreasing the size and transparency of nodes\cite{wilkinson2012grammar}\cite{li2010model}. However, the adjustment usually binds with data, requiring time-consuming customization. Several recent semi-automatic techniques\cite{micallef2017towards}\cite{quadri2020modeling} reduce the workload but are currently focused on single-class scatterplots. For multi-class scatterplots, transparency opens Pandora's box of color blending, which markedly increases the visual complexity and significantly hinders visual tasks, such as cluster identification and class density comparation. Essentially, appearance adjustment cannot eliminate overdraw.

Node dispersion relieves overlap by spreading nodes from their original positions. Many dispersion strategies have been proposed for different layout goals in graph visualization. For example, VPSC\cite{dwyer2005fast-VPSC}, PFS\cite{misue1995layout-PFS}, PFS$'$\cite{hayashi1998layout-PFS'}, PRISM\cite{gansner2010efficient-PRISM}, and FTA\cite{huang2007new-FTA} iteratively approach an ideal dispersion through force-directed or gradient descent techniques. These methods are good at preserving the orthogonal order of input nodes. GTree\cite{nachmanson2016node-GTree} declares a better dispersion on efficiency and shape preservation by growing a minimum spanning tree built on the Delaunay triangulation of nodes. However, GTree reduces space utilization. By contrast, RWordle-L\cite{strobelt2012rolled-RWordle} achieves a compact layout by placing nodes along a spiral curve under the constraint of mutual exclusion. Diamond\cite{meulemans2019efficient-Diamond} develops a stable layout process but sacrifices shape and density preservation. Overall, node dispersion methods usually suffer from three common problems. First, severe distortion frequently occurs, preventing basic visual tasks, such as cluster identification and trend analysis, because these methods do not consider density and shape preservation as mandatory constraints. Second, these methods cannot eliminate overlap due to early termination or falling into local optimum. Third, these methods are computationally inefficient; thus they are only applicable to small data sets. The three problems are exposed by the quantitative evaluation in Section \ref{sec:5.1}.

Subspace mapping methods developed in recent years are the few ones that can eliminate overlap. Typically, these methods first divide visual space into a set of mutually-exclusive subspaces and then map each data point to a subspace by aligning the spatial proximity of two spaces. Subspaces can be generated by isometric grids (DGrid\cite{hilasaca2019overlap-DGrid}, IsoMatch\cite{fried2015isomatch}, Oodanalyzer\cite{chen2020oodanalyzer}), space-filling curves (HaGrid\cite{cutura2021hagrid}), or space-filling treemaps (Nmap \cite{duarte2014nmap}). Compared with node dispersion methods, subspace mapping methods are generally faster and perform better in shape and density preservation. However, the loose coupling of space partition and data distribution leads to severe distortion on shape and density in regions with high density. The distortion can be observed in the qualitative evaluation in Section \ref{sec:5.2}. By contrast, our method is even faster and maintains a high quality data distribution.

\subsection{Hybrid Methods}
Hybrid methods perform data processing and conduct visual encoding or/and view transformation. Bin aggregation\cite{heimerl2018visual-bin}\cite{beilschmidt2019efficient}\cite{luboschik2010new-color-weaving} and contour map\cite{collins2009bubble-contourmap}\cite{mayorga2013splatterplots}\cite{li2019galex} are two representative hybrid methods. In data space, bin aggregation performs data aggregation on data subsets divided by location, while contour map extracts a series of density stairs and corresponding boundaries according to density distribution. Information loss inevitably occurs. In visual space, the two methods discard circular nodes and encode extracted abstract information into visual units with higher level abstraction, such as polygons and paths. Consequently, the one-to-one correspondence between data points and visual units is destroyed. Our method is a hybrid method because it performs data and view transformations. Nevertheless, our method avoids the two aforementioned problems.

\section{Theory Formulation}

\begin{figure}[htb]
  \centering
  \includegraphics[width=0.95\columnwidth]{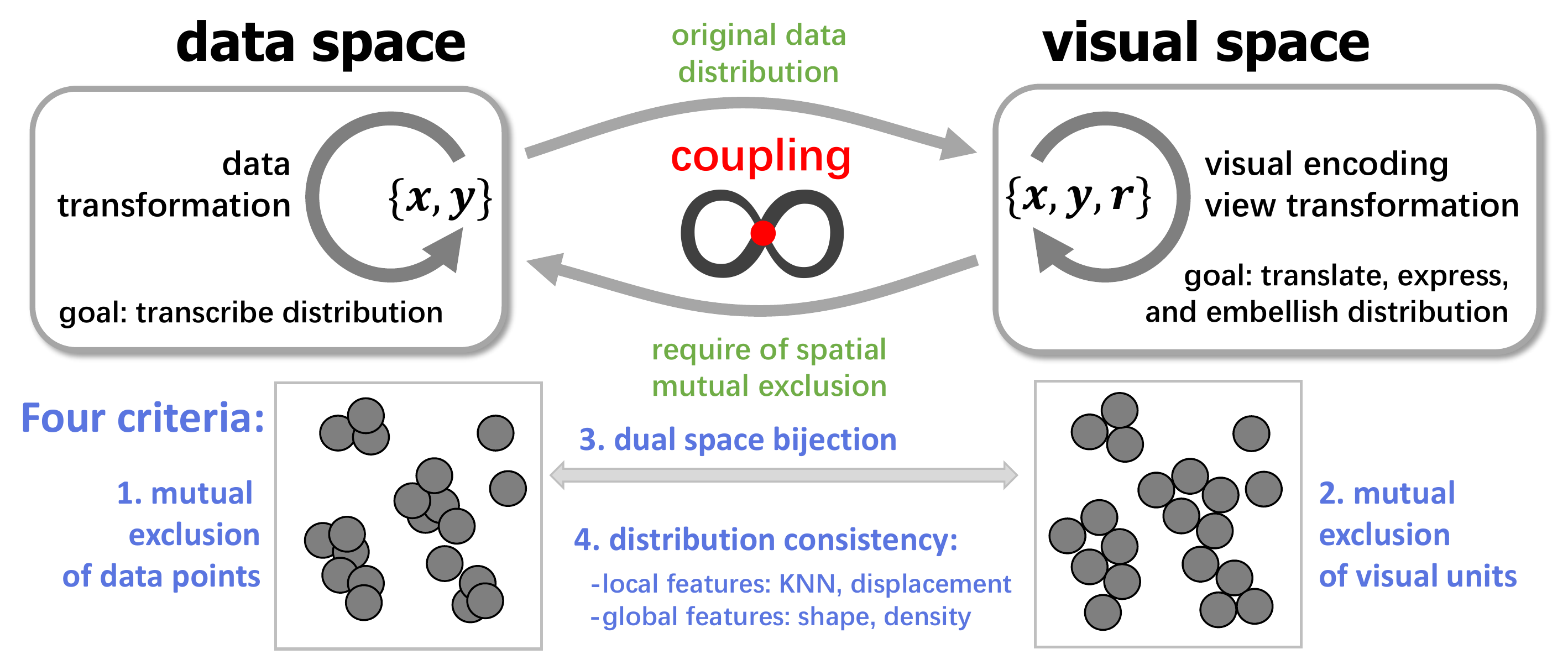}
  \caption{Conceptual illustration of our dual space coupling model.}
  \label{fig:model}
\end{figure}

\subsection{Dual Space Coupling Model}
\label{sec:3.1}

The overdraw problem substantially damages the effectiveness of scatterplot visualization which is reflected by severely hampered common visual tasks of scatterplots, such as cluster identification, shape examination, trend analysis, outlier detection, similar data query, and point value reading\cite{sarikaya2017scatterplots}\cite{yuan2020evaluation-sample}\cite{best2006perceiving}\cite{sedlmair2012taxonomy}\cite{matejka2015dynamic}\cite{doherty2007perception}\cite{gleicher2013perception}. Numerous studies have been conducted to alleviate overdraw, but a complete and detailed analysis of the root cause and a way to eliminate the overlap strictly are still lacking. However, if we re-examine the overdraw and evaluate the existing methods from the perspective of the bilateral relationship between data and visual spaces, then the root cause of the overdraw rises evidently; thus finding a way to solve it theoretically and practically.

The root cause of the overdraw lies in the contradiction between the scale-free and immaterial characteristics of data points and the measurable and corporeal size of visual units. Existing methods have proved that overcoming the conflict (overlaps) inside a single space while ignoring the essential opposition and unity (consistency of data distribution) of the two spaces leads to the non-correspondence and inconsistency between the data points and visual units. The overdraw can be fundamentally solved only by fully considering the unity of opposites between the two spaces, cooperatively solving the contradiction caused by opposites, and maintaining the unity required by visual tasks. Nevertheless, the premise is guaranteeing data integrity.

Assuming the data set in data space is $DS = \{x, \ y\}$ and the visual unit set in visual space is $NS = \{x, \ y, \ r\}$. $x$ and $y$ form the coordinates of a specific object and maybe different in two spaces. $r$ denotes the radius of a specific visual unit which defaults to a circular node. The above considerations can then be formalized into four criteria as follows:
\begin{enumerate}[label=\subscript{C}{{\arabic*}}.]
  \item \textit{Mutual Exclusion of Data Points}: $\forall d_1,\ d_2\in DS,\ \ d_1\cap_{\textbf{\textit{D}}}d_2=\emptyset $
  \item \textit{Mutual Exclusion of Visual Units}: $\forall n_1,\ n_2\in NS,\ \ n_1\cap_{\textbf{\textit{V}}}n_2=\emptyset $
  \item \textit{Data-Visual Space Bijection}: $DS \leftrightarrow NS$
  \item \textit{Data-Visual Space Distribution Consistency}: $F_{\textbf{\textit{V}}}(NS) \sim F_{\textbf{\textit{D}}}(DS)$
\end{enumerate}

$F_{\textbf{\textit{V}}}(DS)$ and $F_{\textbf{\textit{V}}}(NS)$ represent the original and presented data distributions in data and visual spaces, respectively. The first two criteria require overlap-free within a single space. The mutual exclusion of data points is not mandatory because no assumptions should be made regarding the data and it should be satisfied by mandatory mutual exclusion performed in visual space even if nothing is performed in data space. The latter two criteria require the correspondence and consistency between data and visual spaces and are mandated.

Therefore, the goal of a desired scatterplot overdraw solution can be expressed as obtaining a high-quality reconstruction of the data distribution in visual space while ensuring mutual exclusion of data points (optional), mutual exclusion of visual units (mandatory), and data-visual space bijection. This goal can be formalized as:
\begin{equation}
    argmax(similarity(F_{\textbf{\textit{V}}}(NS),\  F_{\textbf{\textit{D}}}(DS))), \ s.t. \ C1, \ C2, \ C3
    \label{formula:1}
\end{equation}

\begin{table}[tbp]
\centering
\resizebox{1\linewidth}{!}{
\begin{tabular}{|c|c|c|c|c|c|}
\hline
\multicolumn{2}{|c|}{\makecell[c]{Categories / Criteria}} & \makecell[c]{C1} & \makecell[c]{C2} & \makecell[c]{C3} & \makecell[c]{C4}
\\ 
\hline
\multirow{3}{*}{\makecell[c]{Data \\ space\\ methods}} & \makecell[c]{data sampling} & \halfcheckmark & \halfcheckmark & \textcolor{red}{\usym{2715}} & \halfcheckmark \\ \cline{2-6}
& \makecell[c]{data aggregation} & \textcolor{red}{\usym{2715}} & \halfcheckmark & \textcolor{red}{\usym{2715}} & \textcolor{red}{\usym{2715}} \\ \cline{2-6}
& \makecell[c]{ jitter } & \textcolor{ao(english)}{\usym{1F5F8}} & \halfcheckmark & \textcolor{ao(english)}{\usym{1F5F8}} & \textcolor{red}{\usym{2715}} 
\\
\hline
\multirow{3}{*}{\makecell[c]{Visual \\space \\methods}} & \makecell[c]{appearance adjustment} & \textcolor{red}{\usym{2715}} & \halfcheckmark & \textcolor{ao(english)}{\usym{1F5F8}} & \textcolor{red}{\usym{2715}} \\ \cline{2-6}
& \makecell[c]{node dispersion} & \textcolor{red}{\usym{2715}} & \halfcheckmark & \textcolor{ao(english)}{\usym{1F5F8}} & \textcolor{red}{\usym{2715}} \\ \cline{2-6}
& \makecell[c]{subspace mapping} & \textcolor{red}{\usym{2715}} & \textcolor{ao(english)}{\usym{1F5F8}} & \textcolor{ao(english)}{\usym{1F5F8}} & \halfcheckmark 
\\
\hline
\multirow{2}{*}{\makecell[c]{Hybrid \\methods}} & \makecell[c]{visual abstraction} & \textcolor{red}{\usym{2715}} & \textcolor{ao(english)}{\usym{1F5F8}} & \textcolor{red}{\usym{2715}} & \textcolor{red}{\usym{2715}} \\ \cline{2-6}
& \makecell[c]{Our method} & \textcolor{ao(english)}{\usym{1F5F8}} & \textcolor{ao(english)}{\usym{1F5F8}} & \textcolor{ao(english)}{\usym{1F5F8}} & \textcolor{ao(english)}{\usym{1F5F8}} 
\\
\hline
\end{tabular}
}
\caption{\label{tab:existing_methods}Comparison with existing representative methods based on the four criteria mentioned in Section \ref{sec:3.1}. \halfcheckmark means ``close to be perfect" or ``is very helpful" but cannot strictly meet the criterion.}
\end{table}

Based on this formulation, we set up a theoretical framework to evaluate overlap removal methods. Table.\ref{tab:existing_methods} presents a summary for comparison between our method and the existing methods. Most existing methods do not meet at least one criterion, which alleviate the overlap but are accompanied by serious negative effects. Then, we propose a dual space coupling model to represent the complex operations, transformations, and design considerations within and between data space and visual space. The model is conceptually illustrated in Fig.\ref{fig:model}. 

The proposed model analytically and theoretically clarifies the design and evaluation principles for feasible solutions. In addition to accommodating the existing techniques, our model introduces a new design space for promising overlap removal algorithms and interaction paradigms and further provides several guidelines for design practice. First, the mutual exclusion of data points cannot guarantee the mutual exclusion of visual units, and the latter can only be achieved by assigning a reasonable size and location for visual units. Furthermore, the latter should not presuppose the former. Second, pursuing the data distribution preservation of visual space is a good starting and driving force to achieve an ideal assignment under the premise of ensuring data lossless and unbiased. Third, visual quality, or visual prominence of visual units, is another important consideration, requiring a trade-off between large visual units and little disruption to distribution preservation. A safe and fast interactive radius configuration tool is necessary to achieve a customized trade-off. The distribution consistency between data and visual spaces is not trivial, and we will discuss metrics of consistency in the next section.

\subsection{Metrics}
\label{sec:3.2}

In this section, we present a measurement framework to measure the ${similarity}$ in Formula \ref{formula:1} quantitatively. The framework is inspired by the visual task of scatterplots and the existing metrics \cite{chen2020NOP-evaluation}\cite{hilasaca2019overlap-DGrid}. This framework comprises an overall metric and four sub-metrics. The overall metric presents a general perspective to measure the similarity of two scatterplots, while each sub-metric has a clear and independent semantics toward visual tasks. Specifically, \textit{displacement minimization} and \textit{KNN preservation} measure the local feature of scatterplots, which are closely related to the visual task of querying and inspecting similar data points. \textit{Density preservation} and \textit{shape preservation} focus on the global feature of scatterplots. The two metrics are designed for visual tasks such as cluster identification, outlier detection, and trend analysis.

Given the original scatterplot $S=\left\{p_i\right\}_N$ and the re-layouted scatterplot $S^\prime=\{p_i^\prime\}_N$, where $p_i$ and $p_i^\prime$ are a pair of corresponding data points in the two scatterplots and $N$ denotes the number of data points. Each metric is briefly described below.

 \textbf{\textit{Displacement minimization}} is the same as \cite{chen2020NOP-evaluation}. We calculate the average of the Euclidean distance between all pairs of $p_i$ and $p_i^\prime$ after scaling $S^\prime$ and $S$ to the same size and aligning their centers. Then the relative displacement, that is, the ratio of the average to the width of the bounding box of $S$, is taken as the final score. The score ranges in $[0, +\infty)$; a small score is superior.

 \textbf{\textit{KNN preservation}} is simply calculated by $\frac{1}{N}\sum_{i=1}^{N}\frac{\left|knn(p_i)\cap{knn}^\prime(p_i^\prime)\right|}{k}$, where $knn(p_i)$ represents the k-nearest neighbors of point $p_i$ in $S$. The score ranges in $[0, 1]$. The $KNN$ is effectively preserved when the score is close to 1.

\noindent\begin{minipage}{0.37\columnwidth}% adapt widths of minipages to your needs
\includegraphics[width=\columnwidth]{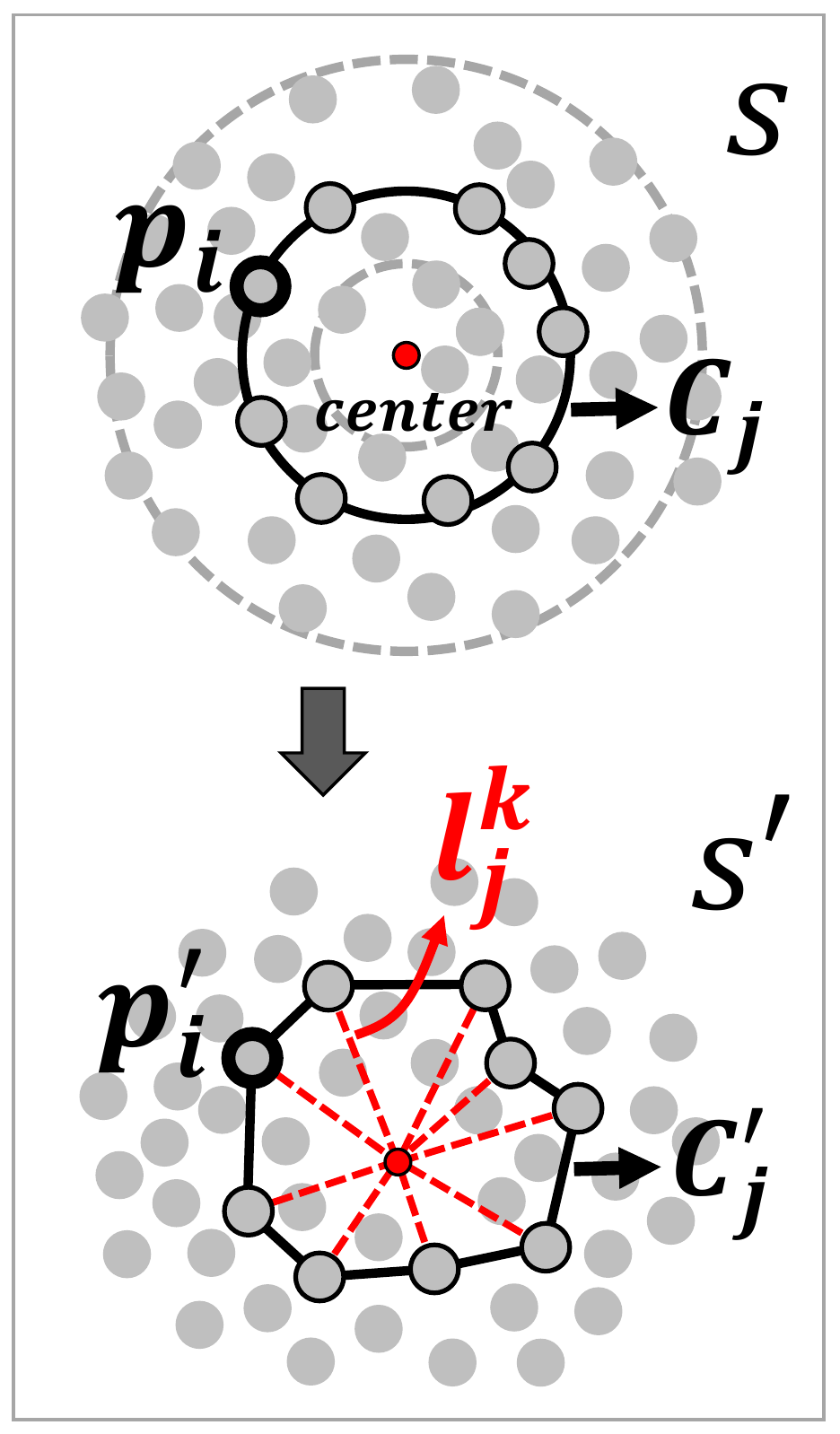}
\label{fig:shape_preservation}
\end{minipage}%
\hfill%
\begin{minipage}{0.6\columnwidth}
 \textbf{\textit{Shape preservation}} extends from \cite{strobelt2012rolled-RWordle} by expanding the measured shape from the outermost contour of the scatterplot to a global scope. We first construct a standard shape in $S$, and then observe whether the shape can be maintained in $S^\prime$. Specifically, as shown in the left, we first construct a group of concentric circles $C_j$ in $S$, and then find a group of points $\{p_i\}^k$ near each circle in $S$. The $k$ denotes the index of circles. Next, we compute the variance of the distance ($l_j^k$) between $\{p_i^\prime\}^k$ and the corresponding center in $S^\prime$ for each circle. Obviously, a small average of the variances facilitates superior shape preservation. The score ranges in $[0, +\infty)$.
\end{minipage}

\noindent\begin{minipage}{0.37\columnwidth}% adapt widths of minipages to your needs
\includegraphics[width=\columnwidth]{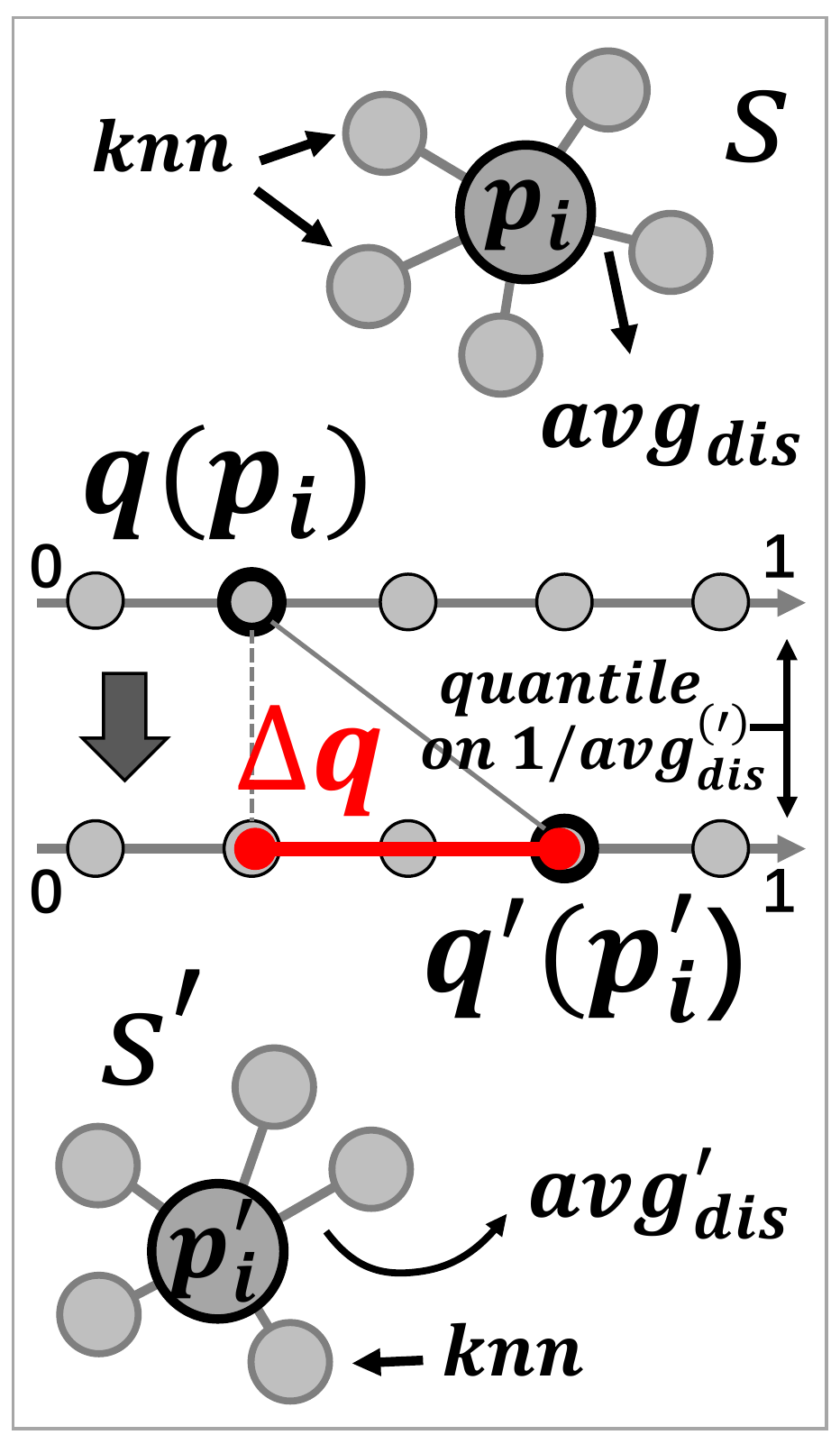}
\label{fig:density_preservation}
\end{minipage}%
\hfill%
\begin{minipage}{0.6\columnwidth}
 \textbf{\textit{Density preservation}} is a new metric that measures the global density preservation by measuring the preservation of relative position of all points in an ordered sequence. Specifically, as shown in the left, for each point $p_i$, we regard the reciprocal of its average distance from its KNN, that is $\frac{1}{{avg}_{dis}}$, as its local density. Then we calculate the quantile $q(p_i)$ of the point $p_i$ in the ordered sequence of all points sorted by the average distance. Therefore, the density preservation score is the average of the difference $\Delta q$ between the quantiles of all paired points in $S$ and $S^\prime$. The score ranges in $[0, 1]$. A score close to 0 is superior.
\end{minipage}

\noindent\begin{minipage}{0.37\columnwidth}% adapt widths of minipages to your needs
\includegraphics[width=\columnwidth]{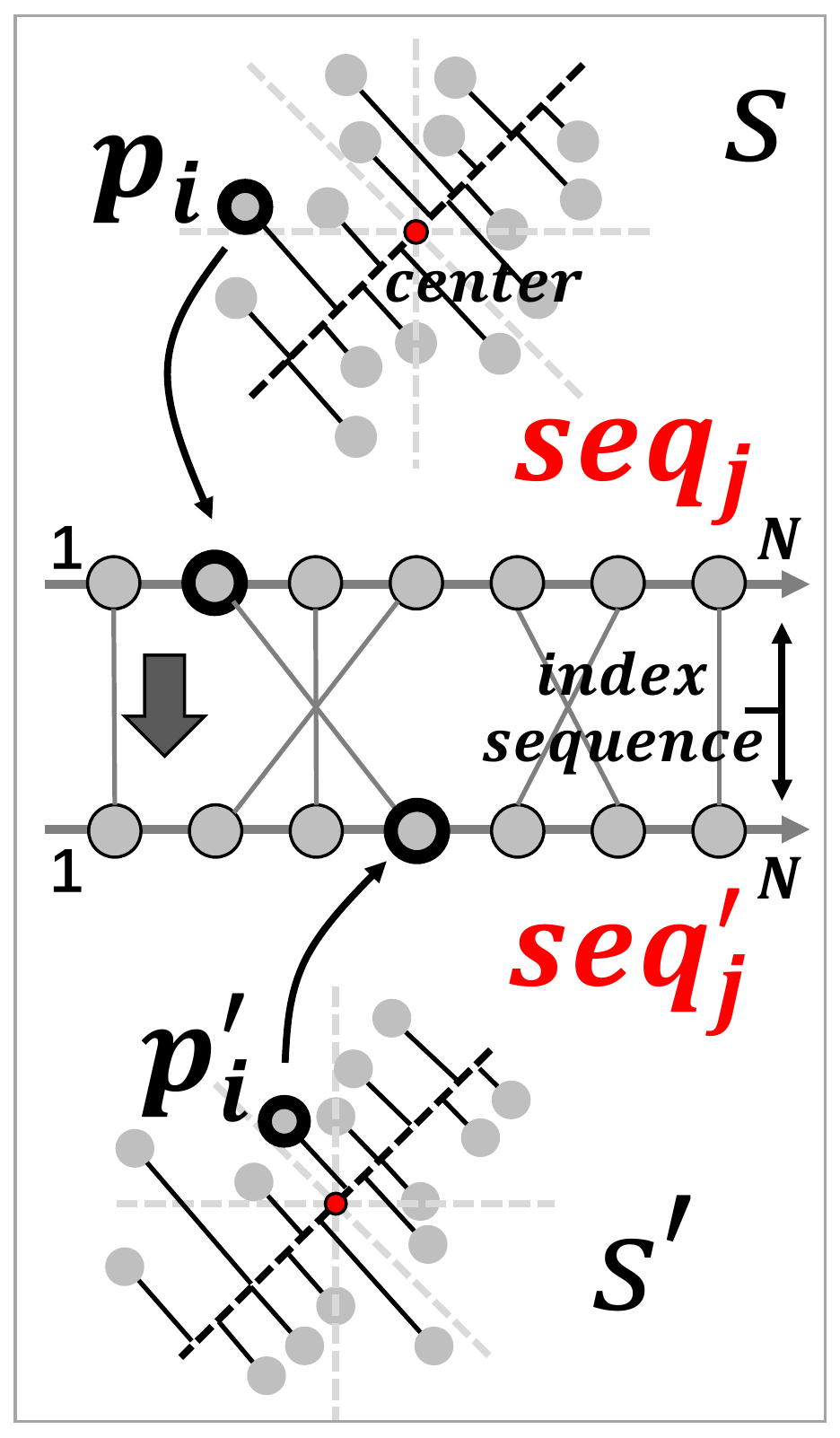}
\label{fig:overall}
\end{minipage}%
\hfill%
\begin{minipage}{0.6\columnwidth}
 \textbf{\textit{Overall similarity}} If $S$ and $S^\prime$ look similar from multiple viewing angles, then they are similar overall. As shown in the left, we first project $S$ and $S^\prime$ on the same set of directions, forming a series of paired ordered index sequences: $\{seq_j\}$ and $\{seq_j^\prime\}$. These directions act as the viewing angles. The Kendall correlation coefficient\cite{EncyMath} is then computed for each pair of the sequences as the similarity between $S$ and $S^\prime$ on the corresponding angle. The final score is the average of all similarities and ranges from $-1$ to $1$. A score close to 1 indicates a high overall similarity between $S$ and $S^\prime$.
\end{minipage}

\section{Methods}
% 方法pipeline
\begin{figure*}[htb]
  \centering
  \includegraphics[width=0.95\textwidth]{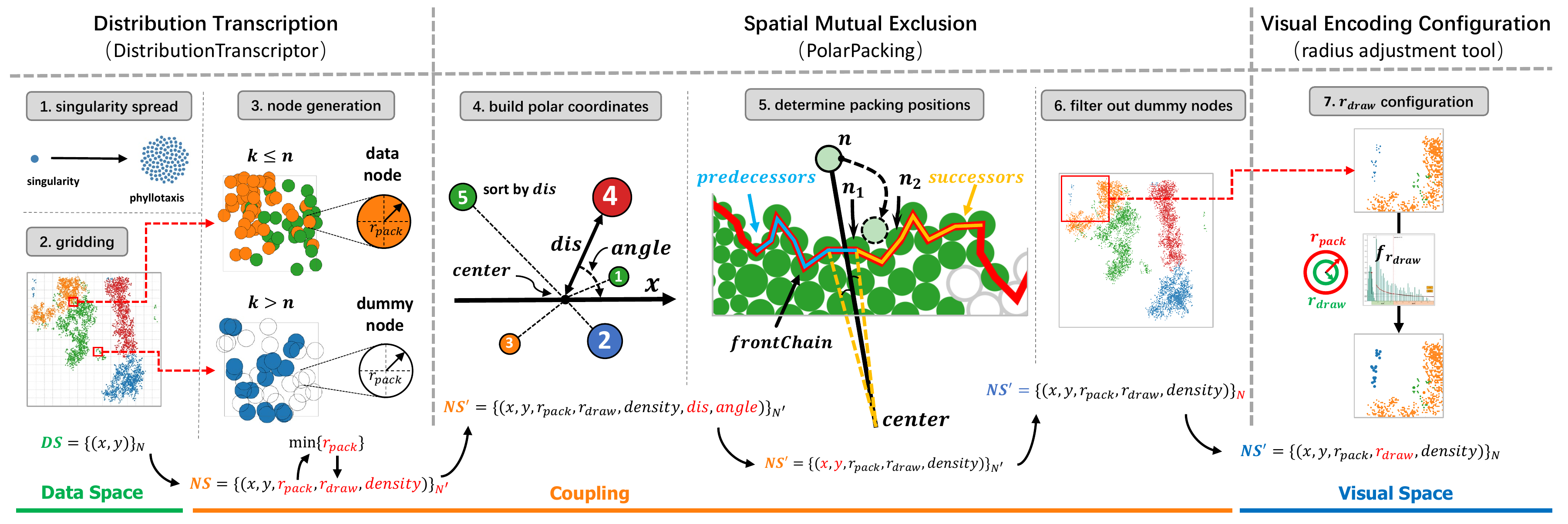}
  \caption{Pipeline of our method. The three components transcribe, translate, and express data distribution from data space to visual space. The algebraic data scheme below presents the evolution of the data flow. The changes between steps are highlighted in red.}
  \label{fig:step1}
\end{figure*}

\begin{figure}[htb]
  \centering
  \includegraphics[width=0.95\columnwidth]{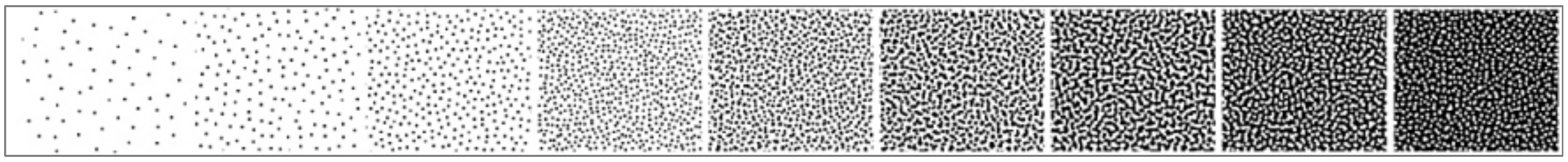}
%   \caption{An example of FM halftoning. The continuous tones are simulated by dots placed at varying frequencies.}
    \caption{Example of FM halftoning.}
  \label{fig:blue_noise}
\end{figure}

On the basis of the dual-space model, we propose a method to build an overlap-free scatterplot and ensure its interaction safety. The core idea is to use a set of closely tangent circles ($NS$) in visual space to imitate the original data distribution in data space ($F_{\textbf{\textit{D}}}(DS)$ in Formula \ref{formula:1}). Some of these circles act as placeholders, filling the blank areas of $F_{\textbf{\textit{D}}}(DS)$. These circles are called dummy nodes and will not be rendered in visual space. The remaining circles, which are called data nodes, are in one-to-one correspondence with the data points. In addition to the radius $r_{pack}$ used for packing, each data node has another radius $r_{draw}$ used for rendering. $r_{pack}$ and $r_{draw}$ are designed to guarantee mutual exclusivity and visibility of nodes, respectively.

The core idea refers to the following three questions: (1) How to generate a set of circles that can record intact $F_{\textbf{\textit{D}}}(DS)$; (2) How to pack these circles sequentially in visual space to express the recorded $F_{\textbf{\textit{D}}}(DS)$ as observable $F_{\textbf{\textit{V}}}(DS)$; (3) How to configure $r_{draw}$ to ensure no overlap occurs (safety) during rendering and interaction. Borrowing concepts from genetics, the first question aims to \textbf{transcribe} the original data distribution from data space to visual space; the second question aims to \textbf{translate} the transcribed distribution from an algebraic and intangible form into a visible and tangible form; the third question aims to \textbf{express} and \textbf{embellish} the distribution in visual space. Fig.\ref{fig:step1} shows the pipeline of our method. A geometry-based data distribution transcription, an efficient spatial mutual exclusion guided view transformation, and an overlap-free oriented visual encoding configuration with an easy-to-use radius adjustment tool are the solutions to the three aforementioned problems. In the following, they will be introduced one by one.

\subsection{Geometry-Based Data Distribution Transcription}

The transcription must maintain the relative position of nodes and the relative density among regions. For the former, we simply inherit the original coordinates of the input data points; for the latter, we borrow the idea of frequency modulation (FM) halftoning, a technique widely used in traditional printing industry\cite{bryngdahl1978halftone}. Fig.\ref{fig:blue_noise} shows an example of FM halftoning. The example simulates continuous-tone imagery through the use of dots that are frequency varying in spacing, thus generating a gradient-like effect. 

We use the similar idea to rebuild varying densities. First, We divide the space into square grids, and then fill each grid with circles. The grid containing $num$ data points will be filled with $max(k, num)$ circles. The packing radius $r_{pack}$ of each circle is equal, given by line 6 of Algorithm \ref{alg:dummy_nodes_creation}. Herein $size$ denotes the size of grids, and $k$ is a threshold representing the minimum number of nodes to be packed in a grid. $size$ and $k$ are all the parameters of our method. $k$ is set to prevent potential distortions in sparse regions. Extra $(k - num)$ dummy nodes acting as placeholders are generated for the grid with less than $k$ data points. The radius of these dummy nodes is also $r_{pack}$. We assign an attribute called $density$ to each data node. The attribute value is given by the ratio of the number of nodes $num$ in the corresponding grid and the maximal $num$ --- ${num}_{max}$. The default value of $r_{draw}$ is $r_{pack}^1$, that is, the minimum $r_{pack}$, obtained in the case of the density is 1. The above settings globally ensure the relative density among regions.

We call a collection of points with the same coordinates thus causing overlaps in data space a singularity. As shown in the upper left corner of Fig.\ref{fig:step1}, we spread each singularity into phyllotaxis to achieve the mutual exclusion of data points. Note that as stated in Section \ref{sec:3.1}, the spread operation is not mandatory. It can be optionally performed before the gridding. However, the spread induces the appearance of singularities as circular agglomerate fogs in visual space which stand out from regular nodes, as if highlighting anomalies.

Algorithm \ref{alg:dummy_nodes_creation} provides the specific steps of the transcription. The input of the algorithm is the original 2D dataset $DS={\{x,\ y\}}_N$, where $N$ is the number of data points. The output is $NS={\{ (x, \ y, \ \textcolor{red}{r_{pack}},  \ \textcolor{red}{r_{draw}}, \ \textcolor{red}{density})\} _{\textcolor{red}{N'}}}$, where $N'$ is the number of nodes to be packed. Hence, $N' - N$ is the number of generated dummy nodes. The red highlights the changes in $NS$ compared with $DS$.

\begin{algorithm}[htb]
  \SetKwInOut{Input}{input}
  \SetKwInOut{Output}{output}
  \let\oldnl\nl% Store \nl in \oldnl
  \newcommand{\nonl}{\renewcommand{\nl}{\let\nl\oldnl}}% Remove line number for one line

  \Input{\\
      \Indm $DS : {\{ (x, \ y)\} _N}  \in R_ + ^2 $      \tcp*[f]{original 2D data}\\ 
        \nonl $ size \in {R_ + } $                    \tcp*[f]{size of grids}\\
        \nonl $k \in {N_ + }$                         \tcp*[f]{minimum number of node in a grid}\\
      }
  
  \Output{\\                                            
          \Indm\nonl$NS : {\{ (x, \ y, \ \textcolor{red}{r_{pack}},  \ \textcolor{red}{r_{draw}}, \ \textcolor{red}{density})\} _{\textcolor{red}{N'}}} \in R_ + ^5$ \tcp*[f]{node set} \\
  }
  \BlankLine
  $DS \gets spreadSingularities(DS)$ \\
  $NS \gets {\emptyset}$\\
    $grids : {\{ (i, \ j, \ subDS)\}}_{m \times n} \gets{gridding(DS, \ size)} $\\
    \For{$grid \ in \ grids$}{
        ${num \gets {\# grid.subDS}}$\\
        ${r_{pack}} \gets{\sqrt {\frac{{siz{e^2}}}{{\pi  \times \max (k, \ num)}}} }$\\
        ${{density} \gets {\frac{num}{{{num_{\max }}}} \in [0,1]}} $\\
        \For(\tcp*[f]{create regular data nodes}){$p \ in \ subDS$}{
            $dataNode \leftarrow (p.x, \ p.y, \ {r_{pack}}, \ density)$\\
            $ NS \leftarrow NS \cup \{ dataNode\} $
        }
                                                      
        \While(\ \tcp*[f]{create dummy nodes}){$num < k$}{                               
          $x \leftarrow grid.j \times size \ + \ random(0, \ size) $\\
          $y \leftarrow grid.i \times size \ + \ random(0, \ size) $\\
          $dummyNode \leftarrow (x, \ y, \ {r_{pack}})$\\
          $NS \leftarrow NS \cup \{ dummyNode\}  $\\
          $num \gets{num \ + \ 1} $\\
        }
        ${n.r_{draw}} \leftarrow \min (\{ n.{r_{pack}}\}), \ n \in NS$   \tcp*[f]{set the default $r_{draw}$}\\ 
        % \For{$ n \ in \ NS$}{$n.{r_{draw}} \leftarrow {r_{draw}}$\\}
        % ${n.r_{draw}} \leftarrow {r_{draw}}, \ n \in NS$ \tcp*[f]{update to default radius}\\ 
    }
    \Return{$NS$}
  \caption{DistributionTranscriptor}
  \label{alg:dummy_nodes_creation}
\end{algorithm}

\subsection{Spatial Mutual Exclusion Guided View Transformation}

Directly rendering the $NS$ from the previous step in visual space is not feasible because the current coordinates and $r_{pack}$ of nodes cannot guarantee the mutual exclusion of nodes. This section presents PolarPacking (Algorithm \ref{alg:PolarPacking}), an algorithm that achieves the required mutual exclusion while maintaining the relative position of nodes. The idea behind the algorithm is to reconstruct the distribution transcribed in $NS$ in polar coordinates.

PolarPacking is modified from CirclePacking\cite{wang2006visualization-circle-packing}. CirclePacking is a layout technique that compactly packs circles without considering their relative positions. PolarPacking must maintain the relative position of the nodes while packing them compactly. As shown in the middle of Fig.\ref{fig:step1}, we first calculate the distance and angle of each node relative to the center of the nodes to convert the relative position encoded in the Cartesian to the polar coordinate system. We then individually pack the nodes in ascending order by their distance. Each node is placed tangent to the outline shaped by the already packed nodes, following its desired angle and packing radius $r_{pack}$. In this way, all nodes, including dummy nodes, are packed from the inside out. During the process, the distance and angle play key roles in maintaining the relative position of nodes, and the $r_{pack}$ and  tangency guarantee strict mutual exclusion. The detailed packing process is presented in Supplementary Material 1.

After filtering out dummy nodes, we get the output of PolarPacking $NS'={\{(\textcolor{red}x,\ \textcolor{red}y,\ {r_{pack}},\ {r_{draw}},\ density)\}_{\textcolor{red}N}}$. The coordinates of nodes in $NS'$ are updated, while the value of other attributes is simply inherited from the previous $NS$. As this point, the goal defined by Formula \ref{formula:1} is achieved if a scatterplot is rendered using $NS'$.

% Theoretically, we can achieve excellent results with PolarPacking on other metrics as long as the quality of the density distribution preserve well.

\begin{algorithm}[htb]
  \SetKwInOut{Input}{input}
  \SetKwInOut{Output}{output}
  
  \newcommand{\nonl}{\renewcommand{\nl}{\let\nl\oldnl}}% Remove line number for one line

  \Input{ \\                                          
          \Indm\nonl$NS : {\{ (x, \ y, \ {r_{pack}},\ {r_{draw}}, \ density)\} _{N^\prime}} \in R_ + ^5$  \tcp*[f]{node set} \\ 
                                                                                      
        %   \nonl$center : (x, \ y) $                       \tcp*[f]{center of the covered rectangle} \\
          \nonl $th \in {N_ + } $                  \tcp*[f]{half the length of subchain}\\                
      }
  
  \Output{\nonl   $ NS' : {\{ (\textcolor{red}{x}, \ \textcolor{red}{y}, \ {r_{pack}}, \ {r_{draw}}, \ density)\} _{\textcolor{red}{N}}} \in R_ + ^5 $ \\}
  \BlankLine
  $NS' \leftarrow \emptyset $\\
  $center \leftarrow$ \ center of $NS$ \\
    \For(\tcp*[f]{build polar coordinates}){$ n \ in \ NS$}{
        $n.dis \gets distance(n, \ center) $\\
        $n.angle \leftarrow angleToxPositive(n, \ center)  $\\
        }
    $ascendingSortByDistance(NS) $\\
                                                      
    % $chain \leftarrow initialize(sortedNS[:3]) $  \tcp*[f]{as in \cite{wang2006visualization-circle-packing}}\\
    %  \tcp*[f]{Make the first three nodes tangent to each other to form the initial fontChain}\\
    $frontChain \leftarrow initialize(NS[:3])$ \tcp*[f]{three tangent circles}\\
    
     \For(\tcp*[f]{pack the remaining nodes}){$n \ in \ NS[3:] $}{  
    $subChain \leftarrow$ \ a subchain of length $2*th$ centered at $n.angle$\\
    $n \leftarrow$ \ the position that has the smallest angle difference to $n.angle$ and tangents to \textbf{only} two nodes on the $subChain$  \\ 
    $ updateFrontChain(n) $ \tcp*[f]{as in \cite{wang2006visualization-circle-packing}}\\
    $ {NS'} \leftarrow {NS'} \cup \{ n\}$\\
    
    %  $n_1 \leftarrow \mathop {\arg \min }\limits_m \left| {n.angle, \ m.angle} \right| \ , \   m \in chain$ \\

    %     $subChain \leftarrow \{predecessor \ of \ n_1\}_{th} \cup \{successor \ of \ n_1\}_{th} $\\
                             
    %     $n_2 \leftarrow findSecondTangentNode(n, n_1), \ {n_2} \in subChain$\\

    %       $ n \leftarrow place(n, \ n_1, \ n_2)$ \tcp*[f]{update n tangent to both $n_1$ and $n_2$}\\
    %       $ updateChain(n) $ \tcp*[f]{as in \cite{wang2006visualization-circle-packing}}\\
    %       $ N{S^'} \leftarrow N{S^'} \cup \{ n\}$\\
        }   
        $filterOutDummyNodes({NS'})$\\
    \Return{${NS'}$}
  \caption{PolarPacking}
  \label{alg:PolarPacking}
\end{algorithm}

\subsection{Overlap-Free Oriented Visual Encoding Configuration}
% rdraw-interaction
\begin{figure}[htb]
  \centering
  \includegraphics[width=0.95\columnwidth]{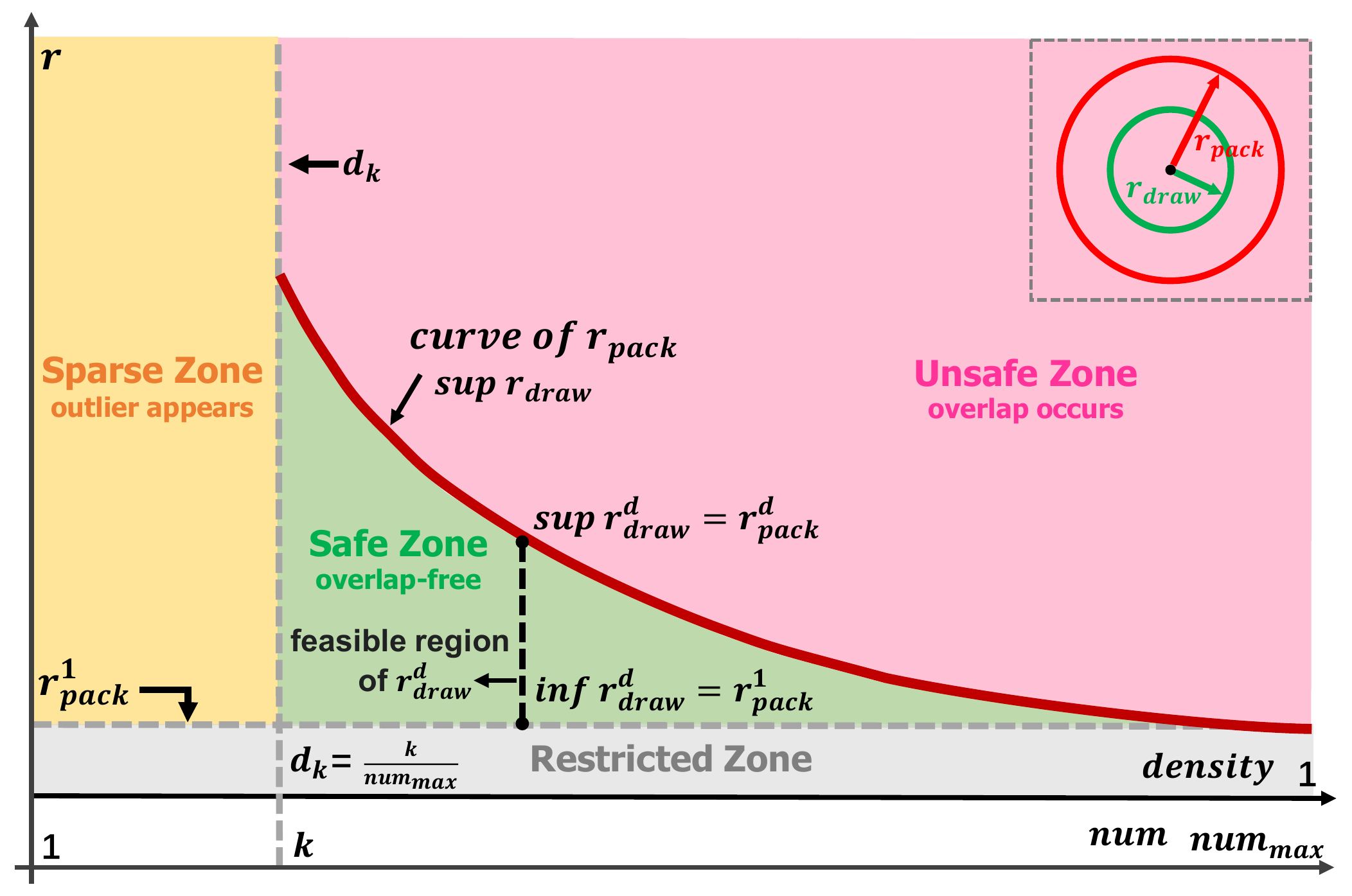}
  \caption{Illustration of our overlap-free visual encoding configuration. The $r$-$density$ space is divided into four zones.}
  \label{fig:r_draw_configuration_model}
\end{figure}
% All nodes are tangent in the output result of PolarPacking algorithm. Tangent nodes cannot express the density features of data space in visual space. 

\begin{figure}[htb]
  \centering
  \includegraphics[width=0.95\columnwidth]{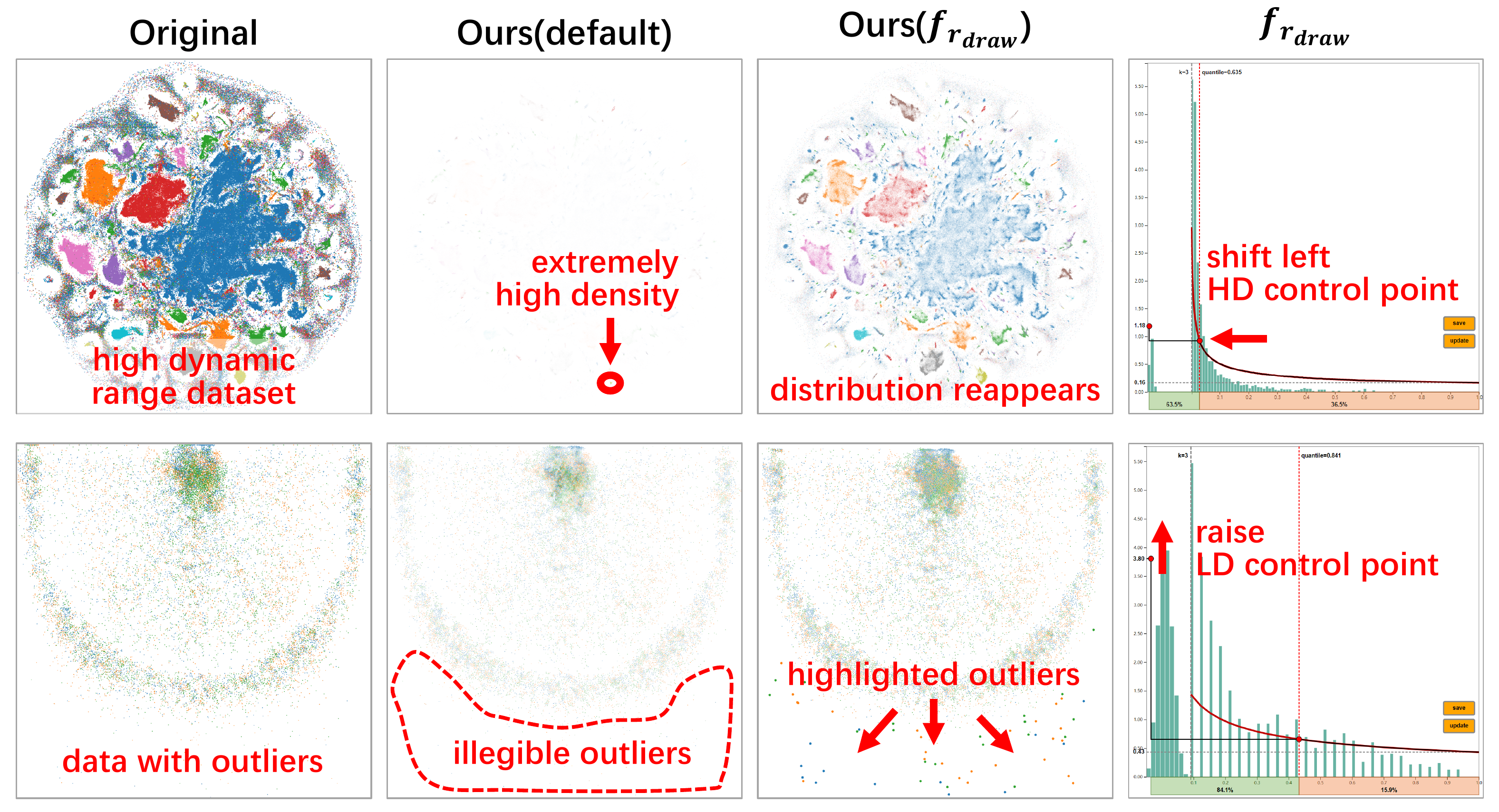}
  \caption{Two instances applying our $f_{r_{draw}}$. The re-appeared data distribution in the first row and the highlighted outliers in the second row respectively demonstrate the necessity and effectiveness of $f_{r_{draw}}$.}
  \label{fig:effect_of_r_adjustment}
\end{figure}

The goal defined by Formula \ref{formula:1} can be achieved using ${NS}^\prime$, but the default $r_{draw}$, that is, $r_{pack}^1$, is inappropriate in some scenarios. For example, consider a high dynamic range dataset (HDR dataset) embedded with regions with extremely high density, which is common for large-scale datasets, its default $r_{pack}^1$ may seriously reduce the visual quality of the corresponding scatterplot. The first row of Fig.\ref{fig:effect_of_r_adjustment} shows an example, where the ``contrast'' of the entire scatterplot is sharply reduced. The reason behind this is that the default $r_{pack}^1$ is usually very small for HDR datasets, leading to an extremely low fill rate of colored informative pixels in most regions. Moreover, outliers are likely to be ignored in this case. In this section, we present a visual encoding configuration model and an interactive radius adjustment tool based on the model to solve the two problems by safely configuring $r_{draw}$ at a expense of distorting density distribution of extraordinary regions.

The configuration model is shown in Fig.\ref{fig:r_draw_configuration_model}. The X-axis and Y-axis represent the $density$ and $r$ of nodes, respectively. The relationship between $density$ and $r_{pack}$ given by line 6 and 7 of Algorithm \ref{alg:dummy_nodes_creation} is a smooth decreasing curve when $num \geq k$. The curve depicts the safe supremum of $r_{draw}$. $r_{pack}^1$ is the infimum of $r_{draw}$. The line $d=d_{k}$ and $r=r_{pack}^1$ divide the entire quadrant into four zones. $d_{k}$ is the density of the grid with $k$ data points. Donate $r_{draw}^d$ and $r_{pack}^d$ are the specific $r_{draw}$ and $r_{pack}$ of regions with density $d$, then $r_{draw}^d$ satisfies ${r_{pack}^1\le r}_{draw}^d\le r_{pack}^d$ in the safe zone; therefore, no overlap occurs. The case is the opposite in the unsafe zone. In the sparse zone, $num > k$ and the $r_{pack}^d$ that acts as the supremum of $r_{draw}^d$ no longer exists. The restricted zone indicates that $r_{draw}^d$ should not be less than $r_{pack}^1$.

\begin{figure}[htb]
  \centering
  \includegraphics[width=0.95\columnwidth]{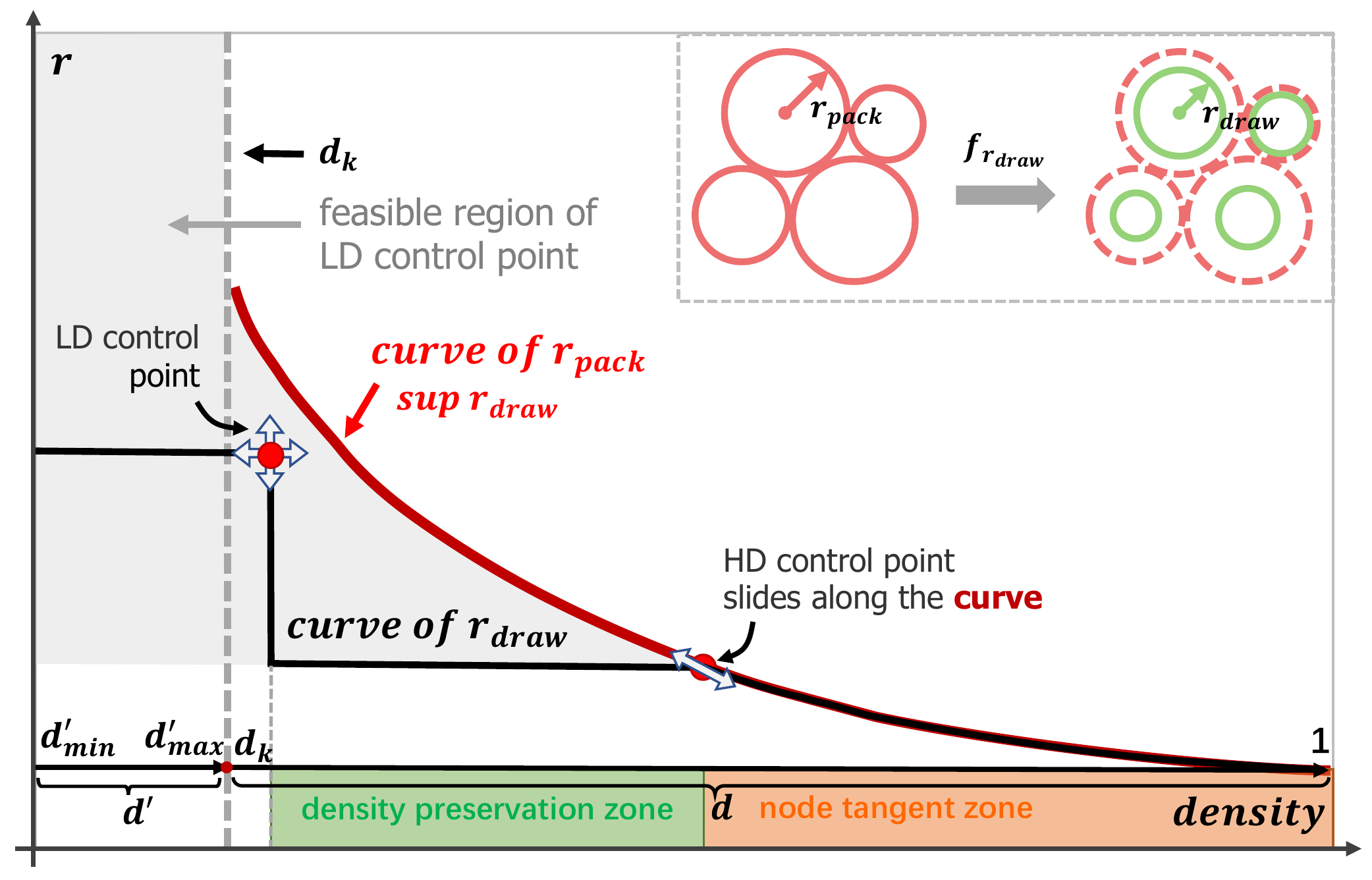}
  \caption{Illustration of our radius adjustment tool $f_{r_{draw}}$. The curve of $r_{draw}$ generated by two flexible points, namely LD and HD control points, determines the rendering radius of nodes.}
  \label{fig:r_draw_configuration_interaction}
\end{figure}

Based on the configuration model, we designed an interactive radius adjustment tool called $f_{r_{draw}}$. As shown in Fig.\ref{fig:r_draw_configuration_interaction}, by moving the high-density (HD) and low-density (LD) control points, we can build a curve of $r_{draw}$ to quickly configure $r_{draw}$ for all nodes. The HD control point can only slide along the curve of $r_{pack}$, while the LD control point can move freely within the gray feasible zone. Therefore, in the right side of the line $d=d_{k}$, the curve of $r_{draw}$ is always under the curve of $r_{pack}$; thus arbitrary configuration here is safe. In the left side, the density of each node is re-assigned to $d^\prime$, which is the reciprocal of the average distance from its five nearest neighbors. Therefore, the range of $density$ is divided into two independent segments: $[d_{min}^\prime,\ d_{max}^\prime]$ and $[d_{k},\ 1]$. Further, the rendering radius $r_{draw}^d$ in the two cases, where the LD control point is located in the left or right side of the line $d=d_{k}$, is given by Formulas \ref{formula:2} and \ref{formula:3}, respectively. In the formulas, $(d_{HD},\ r_{HD})$ and $(d_{LD}^{(\prime)},\ r_{LD})$ represent the coordinates of the two control points. 

The density distribution of the region with density between $d_{LD}^{(\prime)}$ and $d_{HD}$ is preserved, while that of the remaining extraordinary regions is distorted. In the region whose density is larger than $d_{HD}$, nodes are closely tangent to each other, and the fill rate of the colored informative pixel is 100\%. In the region with density less than $d_{LD}^{(\prime)}$, that is, the region where outliers appear, slight overlaps may present due to the absence of a mandatory safe supremum. Notably, the adjustment of $r_{draw}$ using our tool $f_{r_{draw}}$ is independent of the previous transcription and translation steps; thus, the adjustment can be performed in real-time with a WebGL renderer even for large-scale scatterplots.

The last column of Fig.\ref{fig:effect_of_r_adjustment} presents two examples of using $f_{r_{draw}}$. The embedded histogram depicts the node distribution along the density. The second and third columns show the rendering results with the default $r_{pack}^1$ and adjusted $r_{draw}^d$, respectively. The visual quality of the scatterplot of HDR datasets (the first row) and the visual prominence of outliers (the second row) have been markedly improved.

\begin{equation}
{r_{draw}^{d}} = \begin{cases}
      {r_{LD}}   & {d_{min}'}\ \le\ {d'}\ \le\ d_{LD}' \\
      {r_{HD}}   & {d_{LD}'} \ < \ {d'} \ \le \ d_{max}'  \\
      {r_{HD}}   & {d_k}\ \le\ {d}\ <\ {d_{HD}}  \\
      {r_{pack}^d} & {d_{HD}}\ \le\ d\ \le\ 1
      \end{cases} \
      \label{formula:2}
\end{equation}

\begin{equation}
{r_{draw}^{d}} = \begin{cases}
      {r_{LD}}   & {d_{min}'}\ \le\ {d'}\ \le\ d_{max}' \\
      {r_{LD}}   & {d_k}\ \le\ d\ \le\ {d_{LD}} \\
      {r_{HD}}   & {d_{LD}}\ \le\ d\ \le\ {d_{HD}} \\
      {r_{pack}^d} & {d_{HD}}\ <\ d\ \le\ 1
    \end{cases} \
    \label{formula:3}
\end{equation}

\section{Evaluation}
\label{sec:5}
Quantitative evaluation compares the performance of our method with state-of-the-art methods on time cost and the five metrics introduced in Section \ref{sec:3.2} using 50 real-world scatterplots with entirely different distributions. The effectiveness of our method is further demonstrated in qualitative evaluation by showing scatterplots of several representative datasets and the improvements we made in three applications.

% Considering the fact that hybrid methods drop individual nodes and thus change the visual encoding, they aren't our direct competitors.

\subsection{Quantitative Evaluation}\label{sec:5.1}

\textbf{Competing Algorithms, Datasets and Settings} 
Competing algorithms include node dispersion methods, namely PFS$'$\cite{hayashi1998layout-PFS'}, PRISM\cite{gansner2010efficient-PRISM}, GTree\cite{nachmanson2016node-GTree}, and RWordle-L\cite{strobelt2012rolled-RWordle}, and subspace-mapping methods, namely HaGrid\cite{cutura2021hagrid} and DGrid\cite{hilasaca2019overlap-DGrid}. Related algorithms, such as VPSC\cite{dwyer2005fast-VPSC} and Diamond\cite{meulemans2019efficient-Diamond}, are disregarded due to their unacceptable time costs in practice.

\begin{figure}[htb]
  \centering
  \includegraphics[width=0.95\columnwidth]{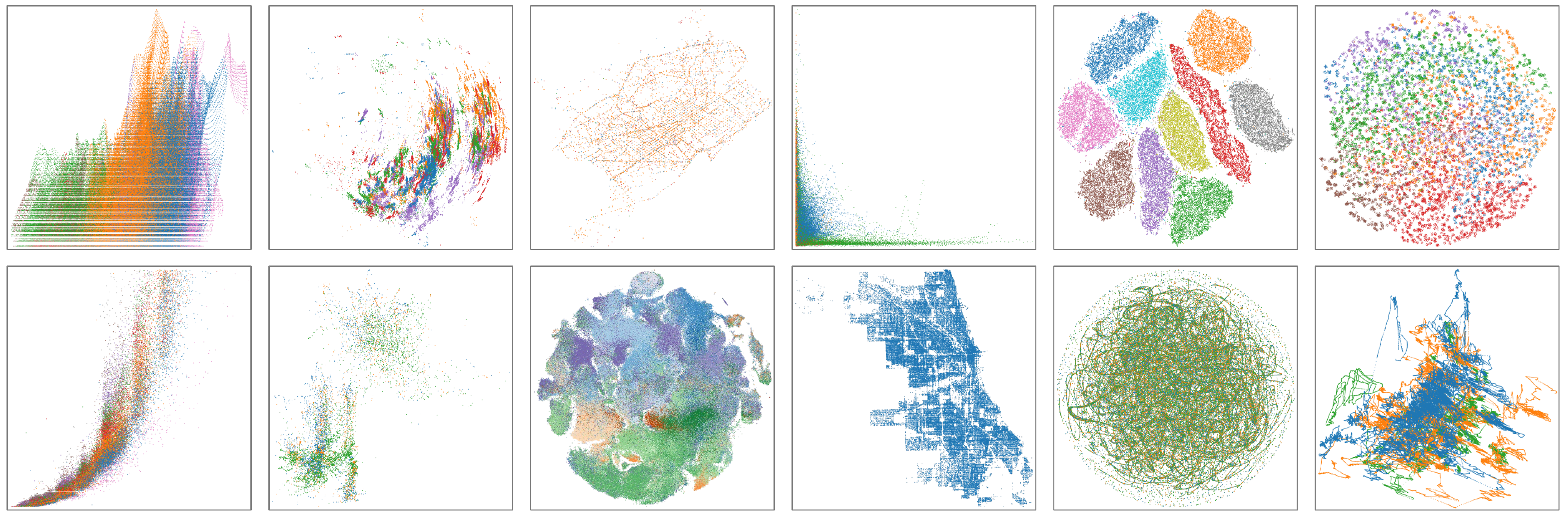}
  \caption{Twelve samples of the collected 50 real-world datasets.}
  \label{fig:datasets_examples_two_rows}
\end{figure}

We collected 50 real-world datasets from \cite{yuan2020evaluation-sample}, \cite{chen2019recursive}, \href{https://archive.ics.uci.edu/ml/index.php}{UCI data repository}, \href{https://networkrepository.com/index.php}{network repository}, and our previous visualization projects. The datasets mainly involve the following four types: regular scatterplots with two semantic axes, projection results of high-dimensional data, coordinates from geographic space, and layout results of large-scale graphs. The number of data points ranges from \num[group-separator={,}]{4177} to \num[group-separator={,}]{928991}. Fifteen of them exceed 100k. The data distribution of these datasets are distinct and some are quite challenging, such as the datasets embedded with extremely high density regions and those with significant features, such as clusters, paths, and textures. Some datasets are shown in Fig.\ref{fig:datasets_examples_two_rows}.

In addition to the full datasets, we created a relatively small data collection, namely Sampled3k, due to the unbearable low computational efficiency of the competing node dispersion methods. This collection is built by randomly sampling 3000 points from each full dataset. The comparison with the node dispersion methods and sub-space mapping methods is performed on the Sampled3k and full datasets, respecitively. All scatterplots are rendered in an $800\times800$ canvas. We take 2.4 as the size of nodes for the node dispersion methods because it strikes a balance between overdraw mitigation and outlier observability.

The implementations of the node dispersion methods are all from \href{https://github.com/agorajs}{\cite{chen2020NOP-evaluation}}, with default parameters. The implementation of HaGrid comes from \href{https://github.com/saehm/hagrid}{\cite{cutura2021hagrid}}. We chose Hilbert curve as the space filling curve and set the depth level of the curve to $l_{min}+1$\, as suggested by the original paper\cite{cutura2021hagrid}. DGrid is implemented by ourselves. Considering the volume of the collected datasets, the number of rows and columns are set to \num[group-separator={,}]{2000} and the size of the convolution kernel is set to $31\times31$, which is a trade-off between computational efficiency and the visibility of local details. The $k$ and $size$ parameters of our method are fixed to 3 and 5, respectively. All the above algorithms are written in JavaScript. The experiment environment is Intel(R) Core(TM) i9-10900K CPU @ 3.70GHz, 64G RAM. The metrics used include time cost and the five metrics introduced in Section \ref{sec:3.2}. The nearest neighbor parameter $K$ used in \textit{KNN Preservation} and \textit{Density Preservation} is set to 10. The number of circles of \textit{Shape Preservation} and the number of directions of \textit{Overall Similarity} are set to 20 and 30, respectively.

\begin{figure*}[htb]
  \centering
  \includegraphics[width=0.95\textwidth]{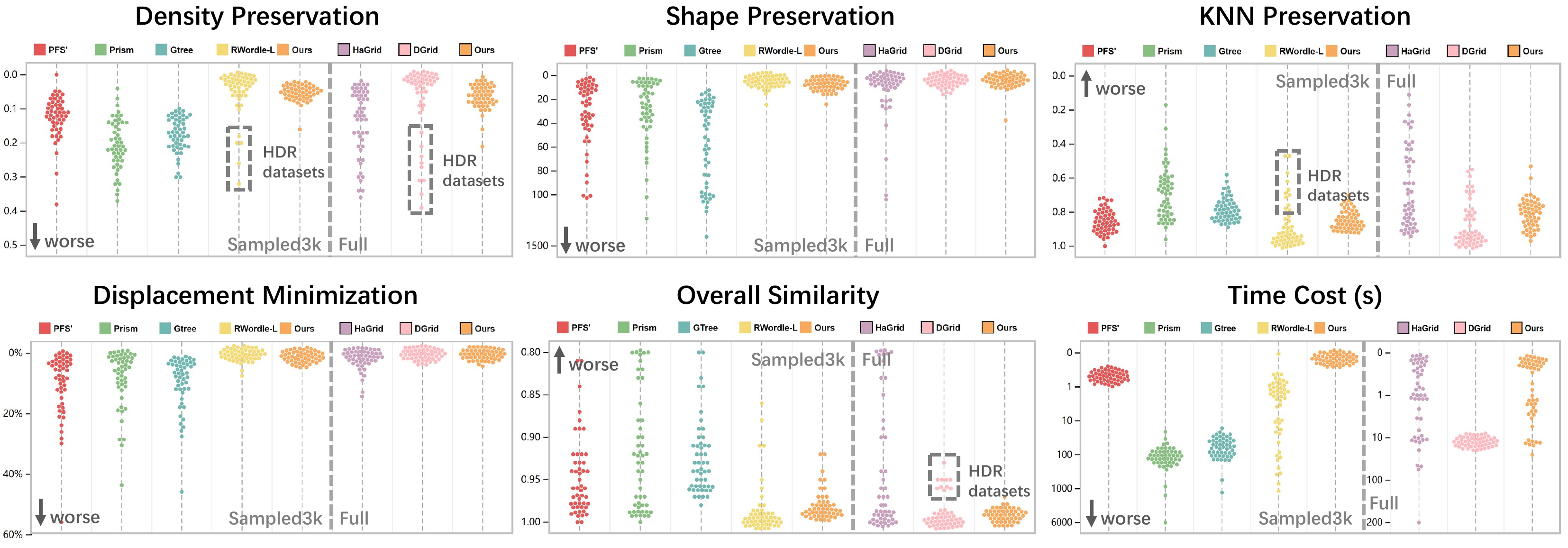}
  \caption{Results of all competing methods on 50 datasets and 6 metrics. Each point in beeswarm represents a dataset. The results show that our method has a considerable advantage in computational efficiency while being the best or comparable to the best on other metrics. In addition, our method shows outstanding adaptability in data distribution. By contrast, other methods often perform poorly on HDR datasets.}
  \label{fig:quantitative_results}
\end{figure*}

\textbf{Results} Fig.\ref{fig:quantitative_results} shows the results of the quantitative evaluation. The left and right parts of each subfigure correspond to the Sampled3k and full datasets, respectively. The results are shown by beeswarm\footnote{https://observablehq.com/@fil/experimental-plot-beeswarm}, wherein each point represents a dataset. First, we note that our method takes excellent scores on all metrics, indicating its effective preservation of semantic features on the whole. Furthermore, for Sampled3k, our method prominently outperforms PFS$'$, PRISM, and GTree in almost all metrics but is equal to or slightly worse than RWordle-L. The average time cost (average on 20 runs) is $1/3.95$ (median $1/3.97$) of the second fastest algorithm PFS$'$ and $1/525.2$ (median $1/17.5$) of the RWordle-L which performs best on \textit{Overall Similarity}. In addition, our method shows better adaptiveness on data distribution compared with RWordle-L, as reflected by its excellent performance on HDR datasets. For the full datasets, our method performs prominently better than HaGrid on \textit{Density Preservation} and \textit{KNN Preservation} and has evident advantages on \textit{Shape Preservation} and \textit{Displacement Minimization} for HDR datasets. Moreover, the average time cost of our method is $1/4.6$ of HaGrid (median $1/2.1$). Compared with DGrid, our method is slightly weaker on \textit{KNN Preservation}, but performs better on \textit{Overall Similarity} and prominently better on \textit{Density preservation} for HDR datasets. The average time cost of our method is $1/47.6$ of DGrid (median $1/46.0$). Generally, our algorithm achieves the best or near the best scores on all metrics compared with the state-of-the-art algorithms. In particular, our method takes great advantage on computational efficiency and presents strong adaptability to HDR datasets. The later will be reconfirmed in qualitative evaluation.

The collected datasets, the implementation of our method and the five metrics, the detailed scores, and the scatterplots created by all algorithms are all available in \href{https://github.com/diyike/scatterplotUnfold}{GitHub}\footnote{https://github.com/diyike/scatterplotUnfold}. The latter two are also presented in Supplementary Material 2. In addition, we implemented a \href{https://diyike.github.io/scatterplotUnfold}{demo}\footnote{https://diyike.github.io/scatterplotUnfold} to interact with the tool $f_{r_{draw}}$ and visually inspect the created scatterplots.

\textbf{Time Complexity} The time complexity of DistributionTranscriptor is $O(N')$, where $N'$ is the number of nodes to be packed. For PolarPacking, the time complexity of sorting nodes is $O(N'logN')$. However, to find the packing position of a given node, $O(\sqrt {N'})$ time is spent to determine a truncated $subChain$, $O(1)$ time to search the final position, and $O(1)$ time to update the $frontChain$. Hence, the overall time complexity of PolarPacking is $O(N'\sqrt {N'})$. Additional details can be found in Supplementary Material 1.

% picture
\begin{figure}[htb]
  \centering
  \includegraphics[width=0.95\columnwidth]{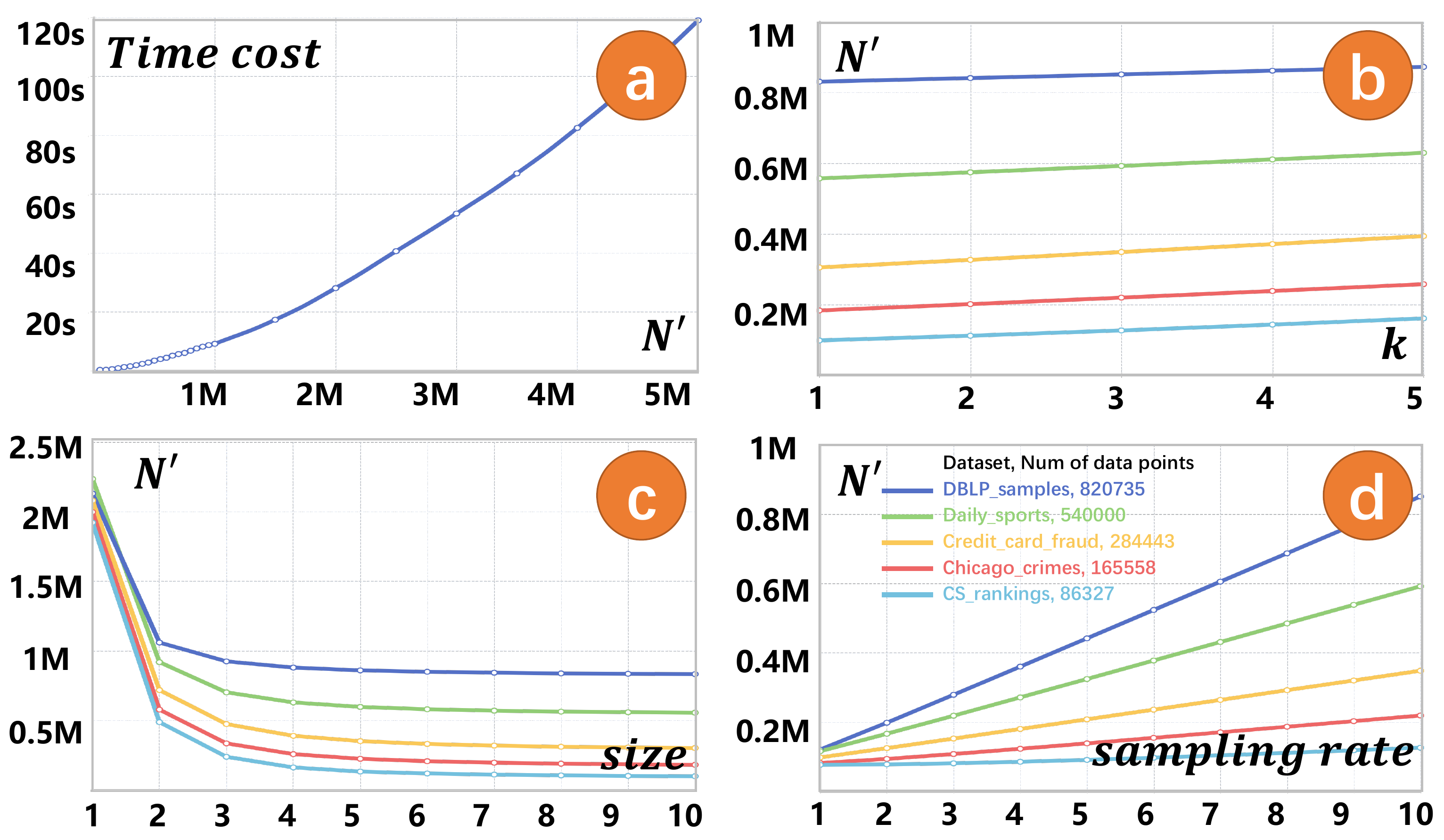}
  \caption{$N'$-time cost curve and $k/size/sampling\_rate$-$N'$ curves. $N'$ is the number of nodes to be packed.}
  \label{fig:effect_of_algorithm_paramters}
\end{figure}

\begin{figure}[htb]
  \centering
  \includegraphics[width=0.95\columnwidth]{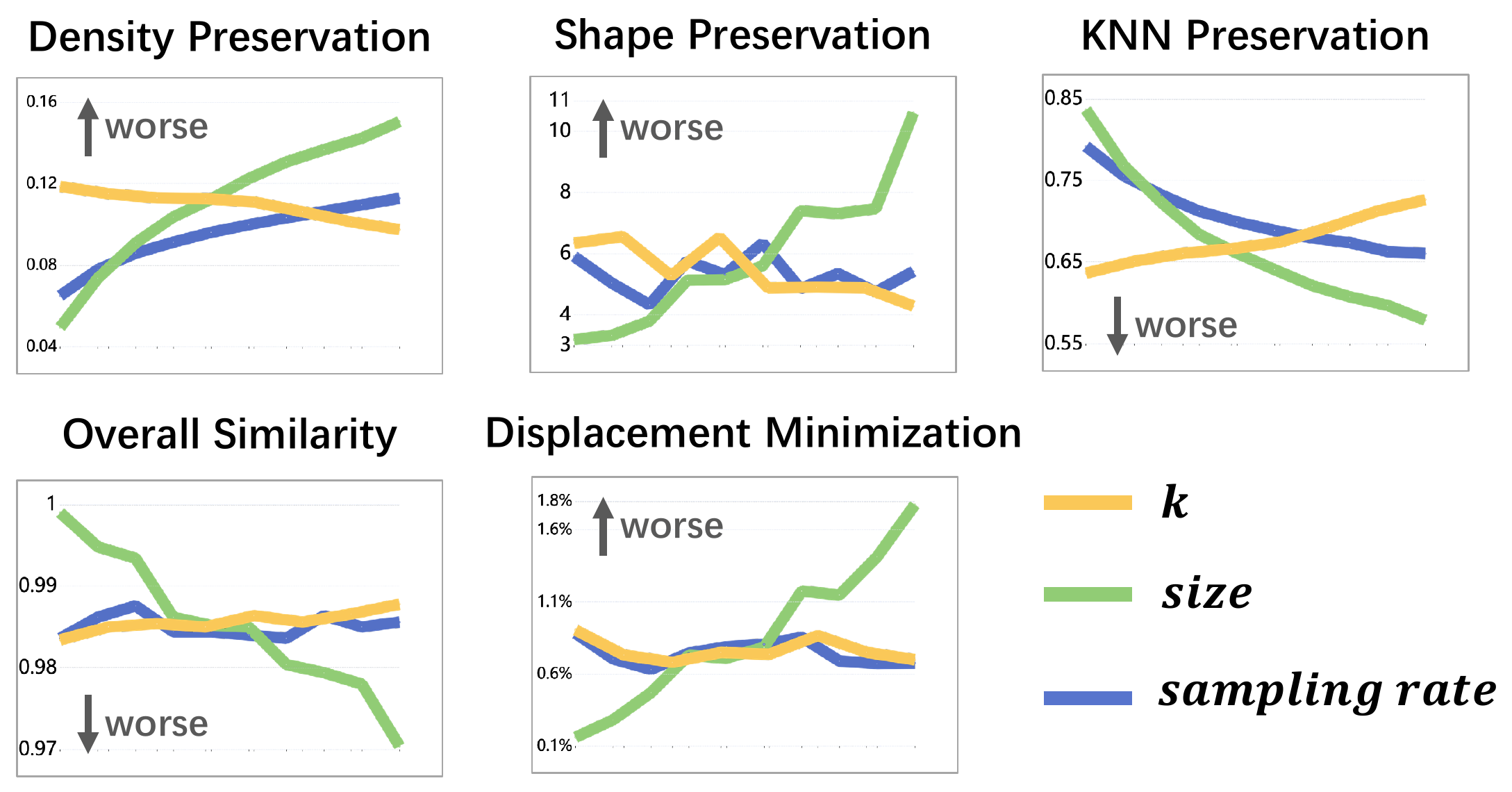}
  \caption{Parameters-metric curves. They demonstrate ours' robustness.}
  \label{fig:impact_of_paramters}
\end{figure}

\textbf{Impact of Parameters} To investigate the impact of parameters $k$, $size$, and $N$ (number of points, equivalent to $sampling\_rate$) on the time cost of PolarPacking in practice, we first studied the relationship between $N'$ and time cost, and then the relationship between the parameters and $N'$. We built a total of 28 simulated datasets whose volume ranges from 5k to 5M. The coordinates and radius of each node are randomly sampled from a unit circle and an interval between 1 and 10, respectively. Fig.\ref{fig:effect_of_algorithm_paramters}\Circled{a} shows the $N'$-Time cost curve. The figure reveals that PolarPacking algorithm is fairly fast, only taking 10s to pack 1 million nodes. Then we selected five representative datasets in volume from the collected datasets, and plotted their $k/size/sampling\_rate$-$N'$ curves (Fig.\ref{fig:effect_of_algorithm_paramters}\Circled{b}\Circled{c}\Circled{d}, respectively). As the interested parameter changes, others are fixed to their defaults (3, 5, and 1 for $k$, $size$, and $sampling\_rate$, respectively). As expected, $N'$ increases as $k$ and $sampling\_rate$ rise and $size$ decreases. The paramter $size$ shows a quadratic-like impact which is considerably larger than the linear-like impact of $k$ and $sampling\_rate$.

% \subsection{Parameter Adjustment}

In addition to time cost, we also investigate the impact of the parameters on the five metrics in our measurement framework. Fig.\ref{fig:impact_of_paramters} presents the results. The range of $size$, $k$, and $sampling\_rate$ is set to [1, 10], [1, 20], and [0, 1], respectively. By mapping these ranges linearly to the same length, the impact curves of all parameters corresponding to the same metric can be aligned in one subfigure. For each curve, the value of the corresponding metric increases along the X-axis, while the others are fixed at their defaults. Noticing the small fluctuation ranges of all metrics on the Y-axis and recalling their intrinsic ranges, we declare that the impact of the parameters on all metrics are controllable, gentle, and reassuring. In other words, our method is fairly robust. Furthermore, we notice that the parameter $size$ has a larger impact than $k$ and $sampling\_ rate$, and all metrics get worse as it raises. This observation is reasonable because $size$, representing the size of grids, determines the global resolution of the captured structures hidden in the data distribution. A small $size$ facilitates the detection and depiction of refined structures by our method. Accordingly, the parameter $k$, representing the minimum number of nodes to be placed in a grid, acts as a local resolution and only shows a notable impact on \textit{Density Preservation} and \textit{KNN Preservation} that depict local features. Interestingly, similar to $k$, $sampling\_rate$ only affects \textit{Density Preservation} and \textit{KNN Preservation}; and the two metrics get worse as it raises.

\begin{figure}[htb]
  \centering
  \includegraphics[width=0.95\columnwidth]{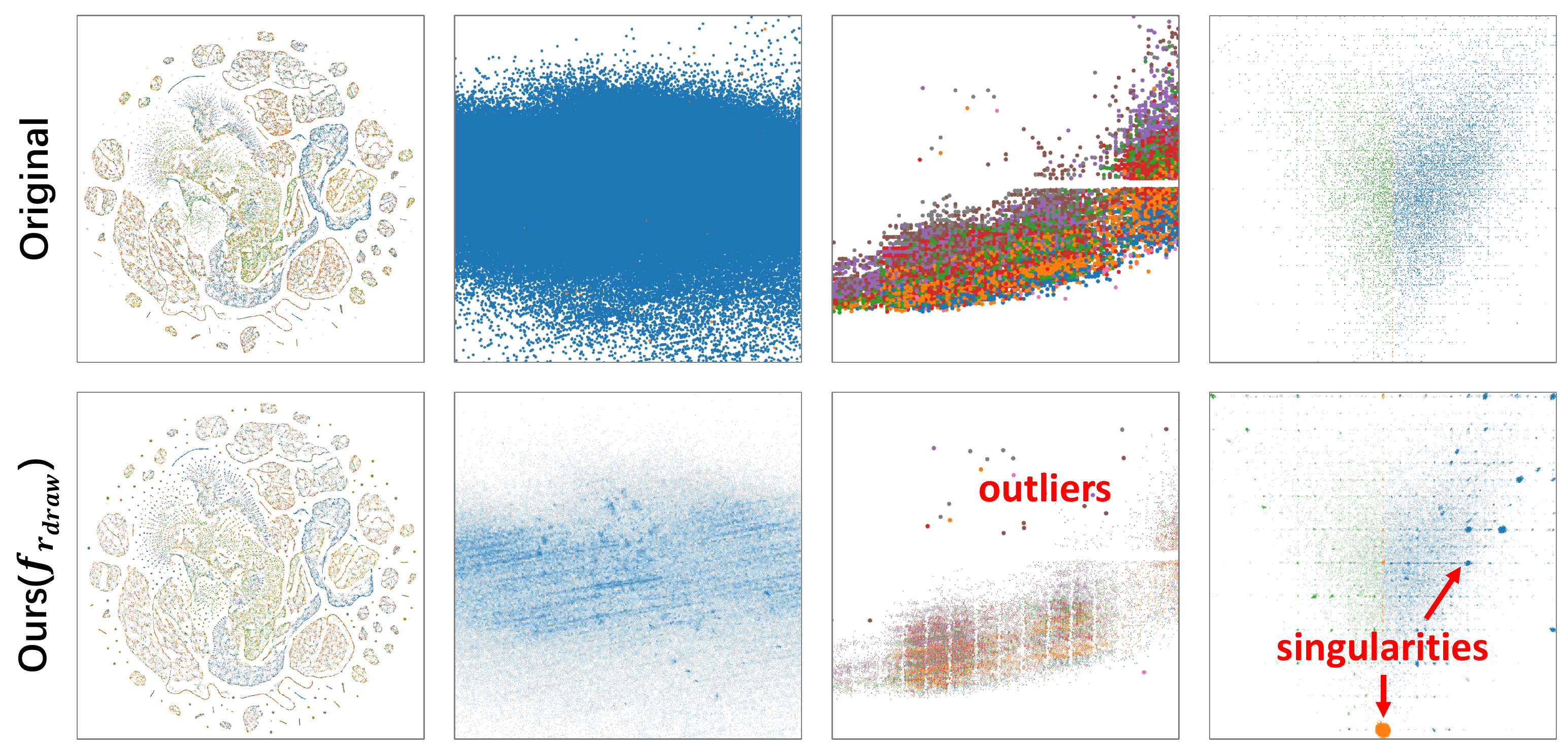}
  \caption{Results of our method on four datasets. Our method can maintain data distribution and reveal details hidden by overdraw.}
  \label{fig:more_examples}
\end{figure}

\subsection{Qualitative Evaluation}
\label{sec:5.2}
% the newly created scatterplots presented in visual space are intuitive and bear the full weight of the prominence of our work.
Compared with quantitative metrics, viewing scatterplots is an intuitive evaluation method. As shown in the first column of Fig.\ref{fig:more_examples}, our method preserve global and local features of the original dataset, such as paths and textures. The subtle textures, equidistant intervals, and circular singularities shown in the last three columns prove the capability of our method to ``reproduce" the intrinsic details covered by overdraw.

\begin{figure}[htb]
  \centering
  \includegraphics[width=0.95\columnwidth]{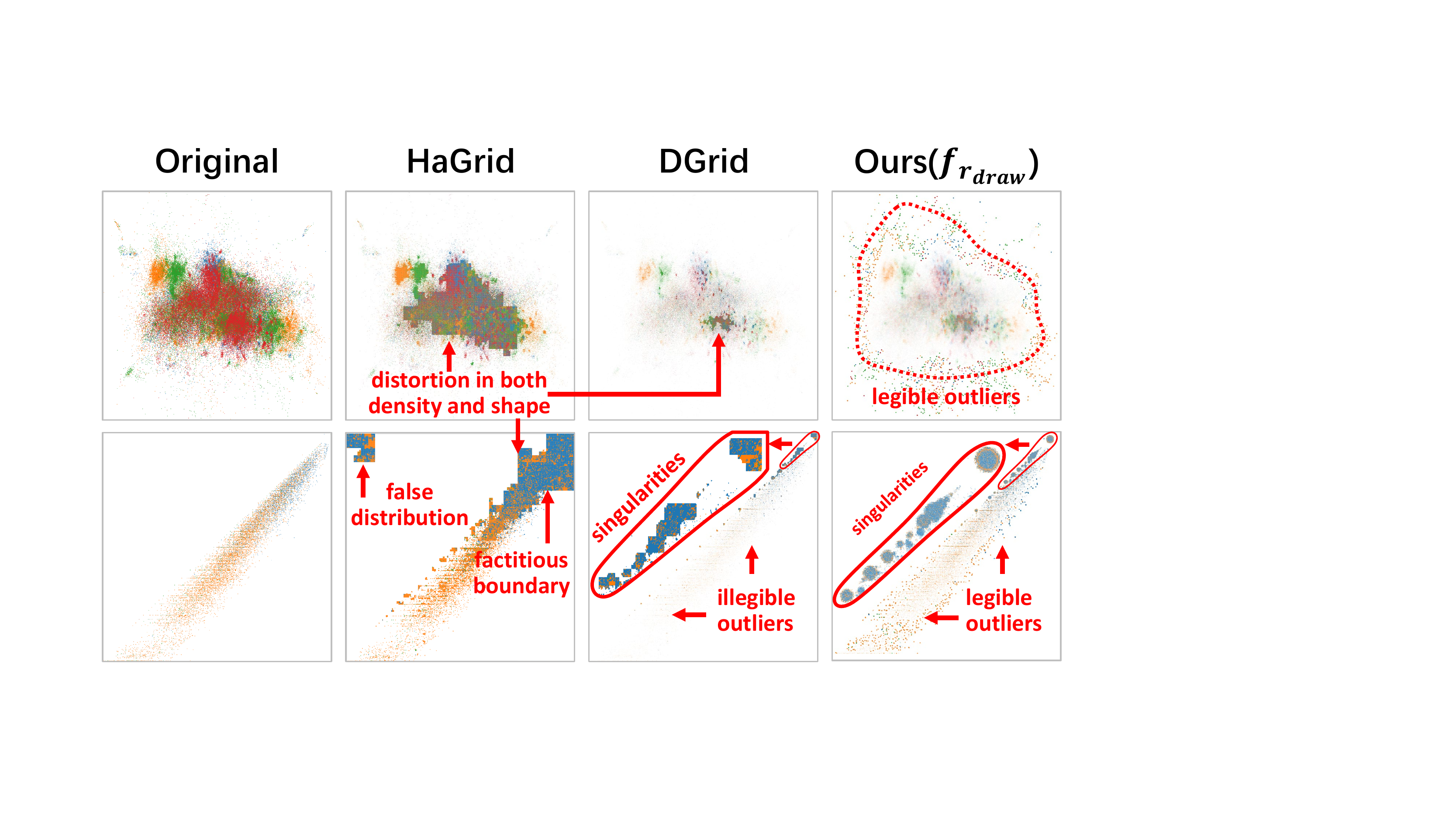}
  \caption{Two examples that demonstrate the crowding issue of HaGrid and DGrid. This issue leads to distortions in shape and density. By contrast, our method accurately depicts the data distribution and reveals more details, such as outliers and singularities.}
  \label{fig:crowding_examples}
\end{figure}

Fig.\ref{fig:teaser} and Fig.\ref{fig:crowding_examples} show the comparison of our method with others. Fig.\ref{fig:teaser} shows that, ostensibly, other methods can reduce the distortion of the density distribution and preserve the overall shape to some extent. However, in fact, their distortions still exist, hiding deceptively, and even new distortions arise. Specifically, neither the sampling nor the transparency adjustment could substantially prevent overlap, and the color blending caused by the latter prevents inspecting details by zooming (Fig.\ref{fig:teaser}\Circled{e}). HaGrid and DGrid locally damage the shape and density preservation. Fig.\ref{fig:teaser} offers the evidence, in which HaGrid and DGrid cannot properly handle regions with extremely high density (Fig.\ref{fig:teaser}\Circled{a}\Circled{b}). The solid blocks cover up all details, including relative density and textures. Moreover, the sharp and straight boundaries are artifacts. We call this issue crowding. It is caused by the failure of the two subspace mapping methods in allocating adequate subspaces for high density regions. By contrast, our method performs well in density and shape preservation (Fig.\ref{fig:teaser}\Circled{c}) and supports seamless zoom to view details (Fig.\ref{fig:teaser}\Circled{d}). Fig.\ref{fig:crowding_examples} presents two additional examples, in which the crowding issue of HaGrid and DGrid leads to misunderstandings. In addition, our radius adjustment tool $f_{r_{draw}}$ can highlight outliers to facilitate observations (Fig.\ref{fig:teaser}\Circled{f}).

\subsection{Applications}
Three applications demonstrate the capabilities of our method in pattern enhancement, interaction improvement, and expandability.

\textbf{Pattern enhancement in semantic space}
% \label{sec:application1}
% 案例1
\begin{figure}[htb]
  \centering
  \includegraphics[width=0.95\columnwidth]{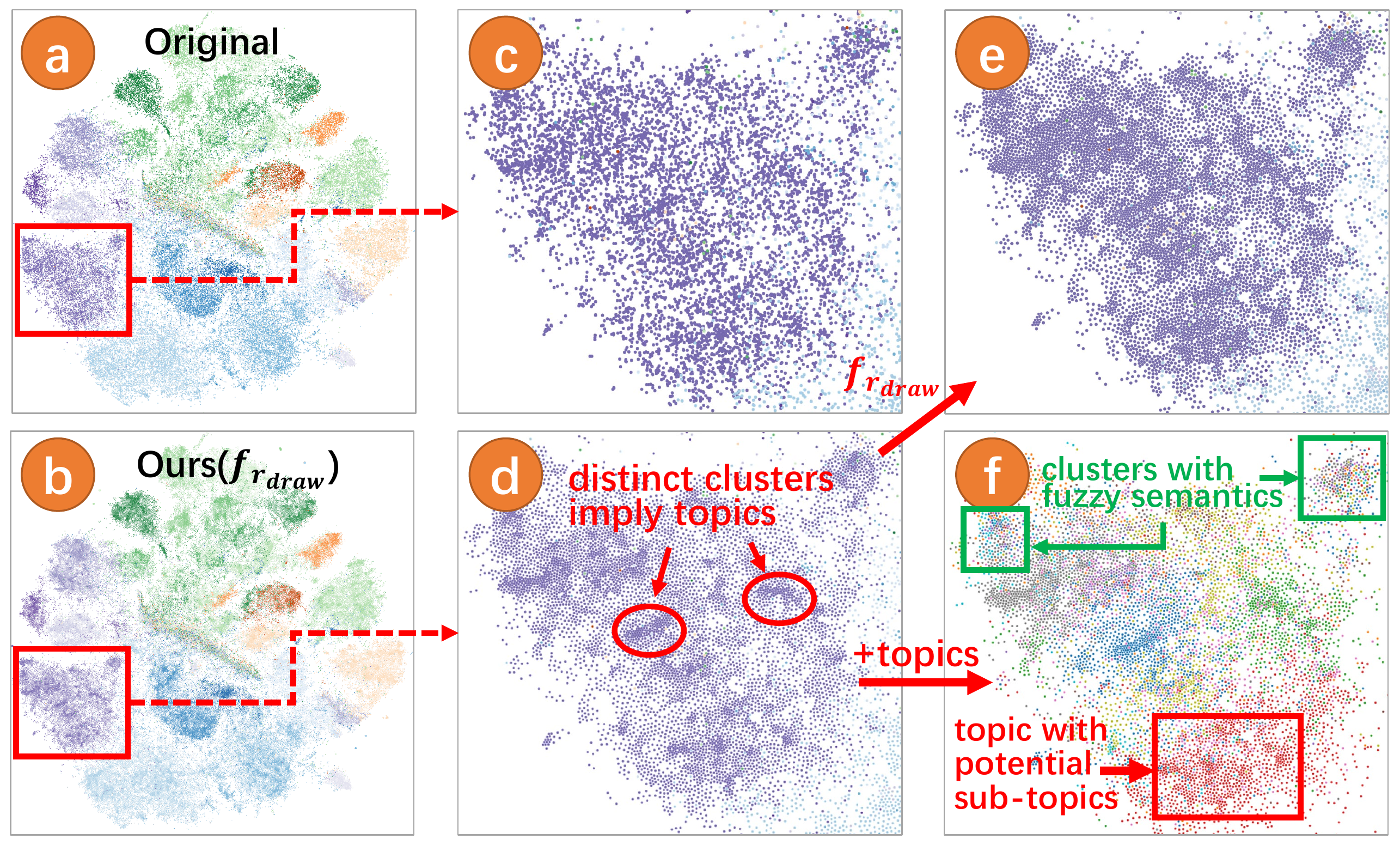}
  \caption{Application of our method to enhance potential patterns in semantic space. Our method (\Circled{b} and \Circled{d}) enhances the visual prominence of the potential clusters in the original scatterplot (\Circled{a} and \Circled{c}). The enhanced clusters and the topics (encoded by colors) uncovered by a topic model complement each other, helping analysts gain a better understanding of the semantic space.}
  \label{fig:application1_pattern_enhancement}
\end{figure}
Fig.\ref{fig:application1_pattern_enhancement}\Circled{a} shows a science map of computer science created by \cite{li2019galex} using 86k scientific literatures. This map acts as the original scatterplot. Fig.\ref{fig:application1_pattern_enhancement}\Circled{b} reveals the new scatterplot created by our method. The differences between \Circled{a} and \Circled{b} and \Circled{c} and \Circled{d} show the power of our method to enhance the visual prominence of potential clusters. Fig.\ref{fig:application1_pattern_enhancement}\Circled{e} shows that our radius adjustment tool $f_{r_{draw}}$ can safely change the intensity and scope of the enhancement without overlaps. These enhanced clusters, like landmarks in cities, quickly attract the visual attention of analysts and elicit their interest, serving as navigators. In Fig.\ref{fig:application1_pattern_enhancement}\Circled{f}, we color each literature by its leading topic which is determined by a probability topic model. The consistency of the spatial distribution between clusters and topics shown in Fig.\ref{fig:application1_pattern_enhancement}\Circled{e} and Fig.\ref{fig:application1_pattern_enhancement}\Circled{f} proves that these clusters can uncover potential topics hidden in the semantic space. More importantly, the semantics provided by the topic model and the spatial structure revealed by the clusters complement each other, jointly promoting the understanding of the semantic space. The reason is two-fold. (1) Clusters remedy the inadequate resolution of topics. As shown in the red box in Fig.\ref{fig:application1_pattern_enhancement}\Circled{f}, the distinct sub-clusters indicate that the red topic can be further divided into sub-topics. (2) Topics help verify whether the clusters have specific, coherent, and understandable semantics. As shown in the green box in Fig.\ref{fig:application1_pattern_enhancement}\Circled{f}, the chaotic distribution of topics implies vague semantics of the focused clusters. We emphasize that all the aforementioned benefits arise from the capability of our method to transfer the correct density distribution from data space to visual space.

\textbf{Interaction improvement in semantic space}
% \label{sec:application2}
% 案例2
\begin{figure}[htb]
  \centering
  \includegraphics[width=0.95\columnwidth]{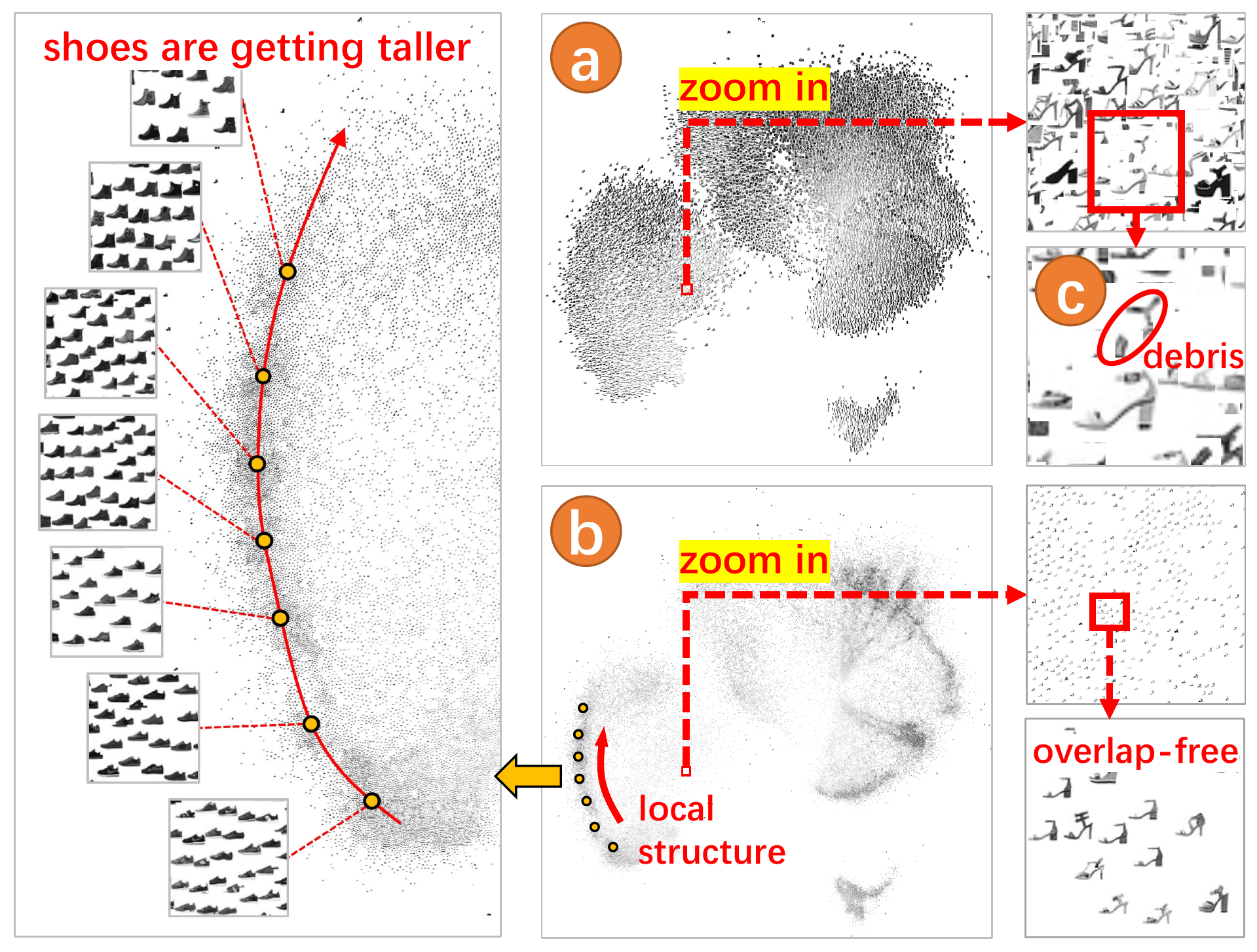}
  \caption{Application of our method to promote exploration efficiency in semantic space. Our overlap-free visualization\Circled{b} avoids image debris and tiring mouse movements, and enables free geometric zooming.}
  \label{fig:application2_interaction_improvement}
\end{figure}
In some scenarios, data points are encoded as snapshots, tiny charts, small multiples, or other glyphs in visual space to support direct observation or statistical analysis of the original data. Fig.\ref{fig:application2_interaction_improvement}\Circled{a} shows a 2D projection of the famous Fashion-MNIST dataset, which includes \num[group-separator={,}]{70000} $28\times28$ gray-scale images. The original visualization suffers a severe overdraw, making the valuable structure invisible. More importantly, overlap markedly reduces the efficiency of interactive exploration. The debris (Fig.\ref{fig:application2_interaction_improvement}\Circled{c}) severely disturbs the reading and even leads to a complete distortion of local semantics. To avoid disturbance, the analyst must place the mouse exactly on the interested image to raise it up and then constantly make movements as the interest changes, which are time-consuming and laborious works. Moreover, the annoying problems cannot be mitigated by a simple geometric zooming. By contrast, without overlaps, our visualization easily reveals the semantic structures (Fig.\ref{fig:application2_interaction_improvement}\Circled{b}) and enables the analyst to grasp the insights hidden in the data quickly by free zoom and pan.

\textbf{Overdraw mitigation of trajectory visualization}
% \label{sec:application3}
% 案例3
\begin{figure}[htb]
  
  \centering
  \includegraphics[width=0.95\columnwidth]{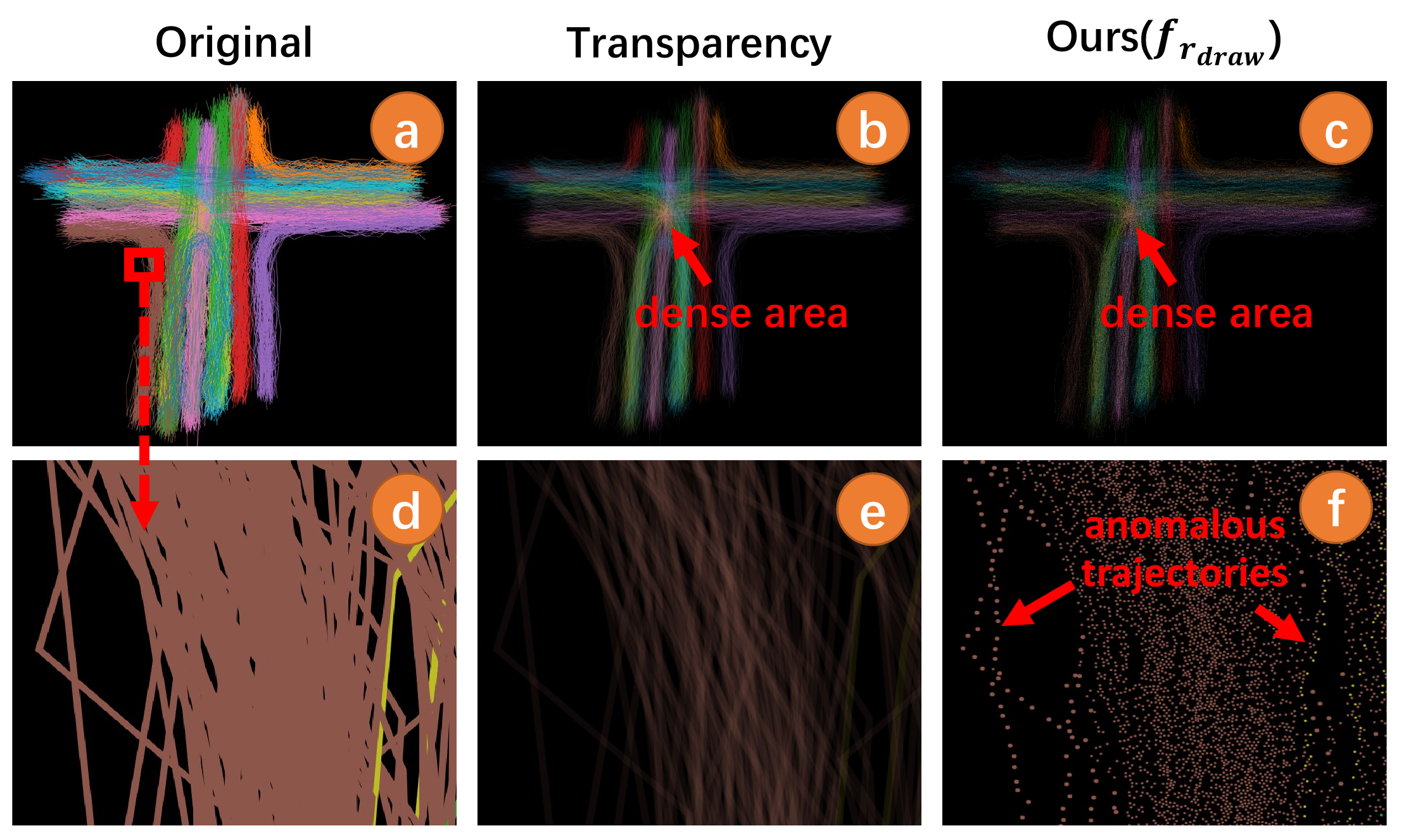}
  \caption{Application of our method in trajectory visualization. In addition to depicting accurate density distribution\Circled{c}, the overlap-free visualization created by our method \Circled{f} has the potential to inspect anomalies that are obscured by regular data (the two yellow trajectories on the right) and are far apart from all others (the two brown trajectories on the left). The two types of anomalies are easily ignored in the original \Circled{d} and low-transparency\Circled{e} visualizations.}
  \label{fig:application3_trajectory_visualization}
\end{figure}
Theoretically, our method can be extended to solve the overdraw of any 2D visualization that can be represented by 2D nodes, such as large-scale time series curves\cite{moritz2018visualizing}, parallel coordinate axis\cite{richer2019hiepaco}\cite{nguyen2017dspcp}, trajectories\cite{zhou2013edge}, and scalar fields\cite{roveri2018correlated}. Here, we take the trajectory visualization as an example. Fig.\ref{fig:application3_trajectory_visualization}\Circled{a} presents an original visualization of vehicle trajectories near a four-lane intersection. The data is taken from the CVPR trajectory clustering dataset\cite{morris2011trajectory}. We sampled massive data points at equal intervals along each trajectory to form the input to our method. Fig.\ref{fig:application3_trajectory_visualization}\Circled{b} and Fig.\ref{fig:application3_trajectory_visualization}\Circled{c} show that both transparency adjustment and our method can reveal regions under great traffic pressure. As shown in Fig.\ref{fig:application3_trajectory_visualization}\Circled{d}, two yellow anomalous trajectories on the right tend to be drowned in massive regular brown trajectories due to the overlap. Unfortunately, reducing transparency does not help, but only increases the likelihood of missing the two brown anomalous trajectories on the left (Fig.\ref{fig:application3_trajectory_visualization}\Circled{d}). By contrast, our method retains both kinds of anomaly (Fig.\ref{fig:application3_trajectory_visualization}\Circled{d}). Though transforming continuous trajectories into discrete points dramatically reduces informative colored pixels, leading to less ``contrast'' of our visualization, the potential of our method in mitigating overdraw of other data types has been successfully demonstrated.

% Moreover, our method can be applied to a wider range of scenes with specific goals by omitting one of the four rule-based constraints. For example, in the first application mentioned above, if the demand for expressing the importance of documents exceeds the demand for expressing the density distribution (e.g., if the analyst expects to intuitively observe the distribution of important documents of each topic), the latter constraint can be sacrificed. The method is to skip algorithm\ref{alg:dummy_nodes_creation} after encoding the importance of the literature with the node radius and run PolarPacking\ref{alg:PolarPacking} directly with that radius. Another example is that analysts can take a sampling algorithm as a preprocessing step of our method to realize the visual abstraction of the scatterplot at the expense of bijection. Compared with direct sampling, this method overcomes the problem that sampling cannot fundamentally avoid overlap.

\section{Conclusion and Future Work}

In this paper, we contribute a dual space coupling model to represent the complex relationship within and between data space and visual space analytically to solve the scatterplot overdraw problem. The proposed model introduces a new design space for promising overlap removal algorithm and interaction paradigm. We also develop an overlap-free scatterplot visualization method on the basis of the model, which shows competitive advantages compared with the state-of-the-art methods.

The algorithms described in this paper are not perfect. The hard partition of space caused by gridding may result in observable regular boundaries, especially when the parameter $size$ is large. A promising solution of this problem is to replace the regular grids with a semantic partition that follows the distribution features, such as superpixels. We leave this idea for future work. Another interesting idea is to extend our algorithms to solve the scalability issues of 3D visualization.

\acknowledgments{
The authors want to thank anonymous reviewers. This work was supported in part by a grant from National Natural Science Foundation of China (\# 62172295).}

\bibliographystyle{abbrv-doi}

\bibliography{template}
\end{document}